\documentclass{article}
\usepackage[a4paper, total={6in, 8in}]{geometry}
\usepackage{setspace}
\usepackage{gensymb}
\usepackage[font=small,labelfont=bf]{caption}
\usepackage{float}
\usepackage{graphicx}

\usepackage{textcomp} 
\usepackage{pifont}   
\usepackage{booktabs}
\usepackage{amssymb,amsthm}
\usepackage{amsmath}
\usepackage{cite}
\usepackage{hyperref}
\hypersetup{hidelinks}
\usepackage{bm}

\begin{document}

\setlength{\parindent}{0em}

\begin{center}
\Large{\textbf{Epitaxy of new layered materials: \\2D chalcogenides and challenges of \\weak van der Waals interactions}}
\end{center}
\doublespacing

\large{Wouter Mortelmans$^{1,2}$, Stefan De Gendt$^{2,3}$, Marc Heyns$^{1,2}$ and Clement Merckling$^{1}$}
\\

\small{$^1$Department of Materials Engineering, KU Leuven, Kasteelpark Arenberg 44, 3001 Leuven, Belgium\\}
\small{$^2$Imec, Kapeldreef 75, 3001 Leuven, Belgium\\}
\small{$^3$Department of Chemistry, KU Leuven, Celestijnenlaan 200f, 3001 Leuven, Belgium\\}

\large{\textbf{ABSTRACT}}
\normalsize{}

The application of new materials in nanotechnology opens new perspectives and enables ground-breaking innovations. Two-dimensional van der Waals materials and more specific, 2D chalcogenides are a promising class of new materials awaiting their usage in the semiconductor industry. However, the integration of van der Waals materials relying on industry-compatible manufacturing processes is still a major challenge. This is currently restricting the application of these new materials to the research laboratories environment only. The large-area and single-crystalline growth of van der Waals materials is one of the most important requirements to meet the challenging demands implied by the semiconductor industry. This review contributes to a more generalized understanding on the integration of van der Waals materials - and in more specific 2D chalcogenides - through the growth process of epitaxy. This, can pursue further the aspiration of large-area, single-crystalline and defect-free epitaxial integration of (quasi) van der Waals homo- and heterostructures into the great world of the semiconductor industry.
\newline


van der Waals epitaxy, quasi van der Waals epitaxy, 2D chalcogenides, molecular beam epitaxy, metalorganic vapor phase epitaxy, chemical vapor deposition

\clearpage

\singlespacing
\tableofcontents

\doublespacing
\setlength{\parskip}{1em}

\section{Introduction: the emerging need for new materials}
\label{section:I}

Materials are the most essential building blocks of everyday life. They shaped our civilization for more than thousands of years and will shape our future society for the many years to come. Materials are constructed from elemental species whose list (and unfortunately occurrence) is finite. The continuous search for ingenious combinations of elemental species to create new materials with new functionalities, has led to breakthrough innovations in many different fields. In the field of nanotechnology, such fundamental materials research is of key importance. This, since nanotechnology concerns the manipulation of matter at the atomic scale and hence directly links with the ultimate material properties. 

To drive the innovation in nanotechnology further, original materials are today being researched from which the lucky few will soon bring new applications to the semiconductor industry. One of these promising materials are the so-called van der Waals (vdW) materials, which are materials having a layered two-dimensional (2D) crystal structure \cite{Miro2014}. As a result, vdW materials are interacting with their environment only through weak vdW interactions in the third dimension and are held together by strong covalent bonding in the 2D plane. Thanks to this fundamentally different crystal structure, they can bring novel functionalities to the semiconductor industry \cite{Geim2013, Kang2015, Zhang2014a, Jena2007}. 

The main driver of the semiconductor industry, already for more than 50 years, is the logic complementary metal oxide semiconductor (CMOS) technology. It relies on the ingenious combination of pMOS (p-type-channel) and nMOS (n-type channel) transistors in integrated circuits (ICs) to dramatically reduce the power consumption of the implemented logic functions \cite{Sah1963}. Its continuous impact in the semiconductor industry makes it the best example to emphasize the innovation driven by materials research and new materials’ integration (examples are: strained silicon-germanium source/drain, high-$\kappa$/metal gate, high-mobility germanium and III-V channel materials,...) \cite{J.Bardeen1950, Arns1998, Moore1998, ITRSa, Heyns2009}. As a result, until today, new materials are continuously being explored to maintain the continuous progress in CMOS technology. Van der Waals materials are such a perfect example of today’s general interest \cite{Kang2015, Zhang2014a, Jena2007}

Moreover, also technologies beyond the CMOS scaling limits are becoming of high interest and are hence being systematically researched, exploring new and yet undisclosed territories beyond classical electron transport such as spintronics, tunnel transistors, topological and low-dimensional devices, optical and quantum computing, etc. \cite{Hutchby2002, Esch2013}. To date, the maturity levels of these technologies are still at the preliminary research stages. Nonetheless, comprehensive and groundbreaking materials research has proven to bring major advancements to these pioneering technologies and will remain essential to enable further developments towards their final integration into the semiconductor industry. Also here, vdW materials are of great interest. 

Besides, developing technologies apart from logic such as flexible electronics, large-area electronics, energy and sensing applications, etc., similarly, highly benefit from materials research and innovation. It is in these perspectives (both logic CMOS and beyond), that materials research focused on vdW materials is predicted to yield promising developments \cite{Akinwande2019, Liu2016a, Schaibley2016, Robinson2018}. 

\section{Van der Waals materials}
\label{section:VDWM}

\subsection{A brief overview}
\label{subsection:ABO}

Pioneering the field, graphene is the most well-known and studied vdW material \cite{Novoselov2004}. It is isolated by K. Novoselov and A. Geim in 2004 and awarded with the Nobel prize in physics in 2010 thanks to its unique transport properties. Graphene is built from carbon atoms with $\text{sp}^{\text{2}}$ hybridized atomic orbitals arranged in a 2D honeycomb lattice. It is known for its superior properties such as the extremely high charge carrier mobility (theoretical limit of above 200 000 $\text{cm}^{\text{2}}$.V.$\text{s}^{\text{-1}}$), the ultralow resistivity ($<$1E-8 $\Omega$.m), the perfect thermal conductivity (1500-2500 W.m.$\text{K}^{\text{-1}}$), superior mechanical strength and stiffness (respectively $\sim$130 GPa and $\sim$1 TPa), the high degree of transparency, etc. \cite{NANOWERK2019}. The main limitation of graphene, however, is the absence of an energy band gap which makes this material challenging to apply in semiconductor devices. 

The absence of an energy band gap in graphene has triggered material scientists to explore the existence of other 2D compounds, which has significantly boosted the research on vdW materials. As a result, besides graphene, to date, also other compounds of the same group IV crystal family are being researched such as silicene, germanene, stanene, etc. \cite{Tokmachev2018}. These more exotic compounds are recently predicted to demonstrate more exotic properties such as strong spin orbit coupling interesting for new topological electronic applications \cite{Ezawa2015}. Beside the graphene-like crystal family, also other vdW crystal families show topological properties such as for example compounds from the group $\text{V}_{\text{2}}$$\text{VI}_{\text{3}}$ family like $\text{Bi}_{\text{2}}$$\text{Se}_{\text{3}}$ and $\text{Sb}_{\text{2}}$$\text{Te}_{\text{3}}$, as well as their alloys $\text{(Bi,Sb)}_{\text{2}}$$\text{(Se,Te)}_{\text{3}}$ \cite{Moore2010, Liu2019a}. Moreover, phase-change vdW materials and transition metal dichalcogenides (TMDs) such as for example GeSbTe compounds and Mo-/W-sulfides or -selenides, respectively, are currently also gaining more and more attention thanks to their interesting application potential in logic, memory and energy storage devices \cite{Robinson2018, Sorkin2014, Yoo2019, Sivan2019, Radisavljevic2011a}. In addition, freestanding "2D" oxides like $\text{BiFeO}_{\text{3}}$ \cite{Ji2019} and 2D magnetic vdW materials like $\text{Fe}_{\text{3}}$$\text{GeTe}_{\text{2}}$ \cite{Gibertini2019} are being demonstrated to hold promise for future nanoelectronic applications. 

Furthermore, the stacking of vdW materials to vdW heterostructures further enlarges the potential of vdW materials \cite{Geim2013, Novoselov2016, Liang2019}. This, since at once exceptional properties can be precisely engineered by combining unique vdW layers. These exceptional properties include for example superconductivity \cite{Geim2013, Zhang2018c}, ferro- and electromagnetics \cite{Arai2015, Behera2019, Klein2018, Zhong2017}, favorable tunneling junctions \cite{Arai2015, Klein2018}, ... Moreover, vdW heterostructures have also been effectively applied in field effect transistors \cite{Liang2019, Roy2014}, memory \cite{Li2019, Tran2019}, photovoltaic and light emitting devices \cite{Novoselov2016}. 

For all of the above future nano- and optoelectronic applications, one of the most appealing types of vdW materials, are the layered 2D chalcogenides \cite{Kolobov}. 

\subsection{2D chalcogenides}

\begin{figure}
  \centering
  \medskip
  \includegraphics[width=1\textwidth]{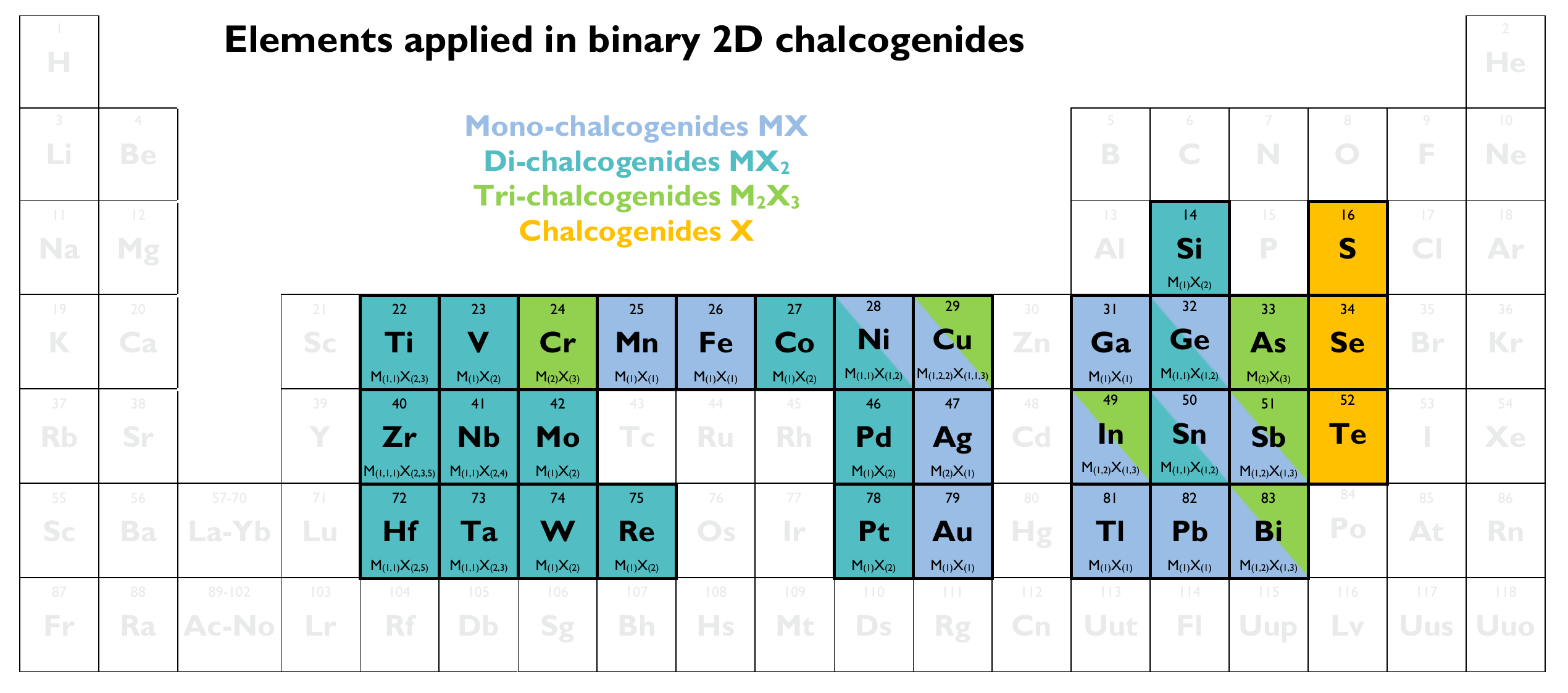}
  \caption[1]{Periodic table of elements highlighting the chemical elements that are currently applied in commercially available binary 2D chalcogenides. These 2D chalcogenides exist in the chemical compounds of mono-chalcogenides in the form of MX (blue), di-chalcogenides in the form of $\text{MX}_{\text{2}}$ (aqua) and tri-chalcogenides in the form of $\text{M}_{\text{2}}$$\text{X}_{\text{3}}$ (green). The chalcogenide elements (X) are colored yellow. The available 2D chalcogenide compounds for each element are presented in the bottom of the cells. \cite{2dsemiconductors, hqgraphene}}
  \label{figure:1.1}
\end{figure}

An overview of the elements that are applied as building blocks in commercially available binary 2D chalcogenide materials is presented in Figure \ref{figure:1.1} \cite{2dsemiconductors, hqgraphene}. These binary compounds include mono-chalcogenides (MX, blue), transition metal dichalcogenides ($\text{MX}_{\text{2}}$, aqua), and tri-chalcogenides in the chemical form of ($\text{M}_{\text{2}}$$\text{X}_{\text{3}}$, green). The chalcogen elements (S, Se and Te) are highlighted in yellow. The broad range of elements used, as well as the wide range of possible binary compound combinations, make 2D chalcogenides a versatile class of vdW materials, covering a broad variety of interesting properties for nano- and optoelectronic applications. Moreover, the continuously expanding library of elements used in 2D chalcogenides, gives rise to new functionalities which significantly enlarges the potential of these materials yielding promising new applications \cite{Liu2016a}. To date, hence many new 2D chalcogenide materials are under investigation for their integration into the semiconductor industry. In this review, the focus will mainly be set on epitaxial di- and tri-chalcogenides (respectively, $\text{MX}_{\text{2}}$ and $\text{M}_{\text{2}}$$\text{X}_{\text{3}}$).

\subsubsection{Properties} 
\label{subsubsection:P} 

The most attractive property for MX and $\text{MX}_{\text{2}}$ compounds is the presence of the energy band gap making these compounds semiconducting. This property attracted researchers’ interests and has significantly boosted the research and development of these vdW compounds. Moreover, the range of the energy band gaps covered by these compounds is in the very desirable window for nanoelectronic applications (between $\sim$0.5-2.0 eV). Even more, in some specific cases, the band gap nature is direct (instead of indirect for silicon), which additionally attracts lots of interests for optoelectronic applications \cite{Wang2012}.

Besides the band gap, these compounds also exhibit other interesting properties for advanced semiconductor devices and applications. These are for example the different band offsets interesting for tunnel devices \cite{Zhao2018a}. This, since sharp staggered and broken gap band structures can hence be precisely engineered, predicted to facilitate optimal tunnel performances \cite{Koswatta2010}. Moreover, the charge carrier mobilities in the order of 100-1000 $\text{cm}^{\text{2}}$.V.$\text{s}^{\text{-1}}$ at single-layer thicknesses are tempting and outperforming ultra-scaled silicon \cite{Robinson2018, Li2019b}. This combined with the low dielectric constants (generally lower than silicon ($\kappa$=12)) results in superb electrostatic control and significantly reduced short-channel effects \cite{Laturia2018}. Furthermore, exotic compounds such as $\text{SnSe}_{\text{2}}$, $\text{ZrTe}_{\text{2}}$, $\text{MoTe}_{\text{2}}$, $\text{VSe}_{\text{2}}$, $\text{TiTe}_{\text{2}}$, etc. exhibit exotic properties in 2D layers such as superconductivity \cite{Hsu2017, Zhang2018c}, massless Dirac fermions \cite{Tsipas2018}, Weyl fermions \cite{Belopolski2016, Tsipas2018a, Belopolski2016a}, charge-density waves \cite{Feng2018, Fragkos2019}, .... The application potentials of these MX and $\text{MX}_{\text{2}}$ vdW materials and their vdW heterostructures are therefore unlimited.

The most appealing property of $\text{M}_{\text{2}}$$\text{X}_{\text{3}}$ compounds is its topology. This is resulting from their characteristic electronic band structure that demonstrates non-trivial band topology, which gives rise to robust, spin-polarized electronic states with linear energy–momentum dispersion at the surfaces of these 2D chalcogenide materials \cite{Liu2019a}. The precise control and accurate detection of the topological states, will make 2D $\text{M}_{\text{2}}$$\text{X}_{\text{3}}$ chalcogenides very useful in new electronic devices, making 2D topological insulators a promising class of future materials for novel spintronics and topological applications \cite{Liu2019a, Moore2010}.

\subsubsection{Crystal structures}
\label{subsubsection:CS}

\begin{figure}[!t]
  \centering
  \medskip
  \includegraphics[width=1\textwidth]{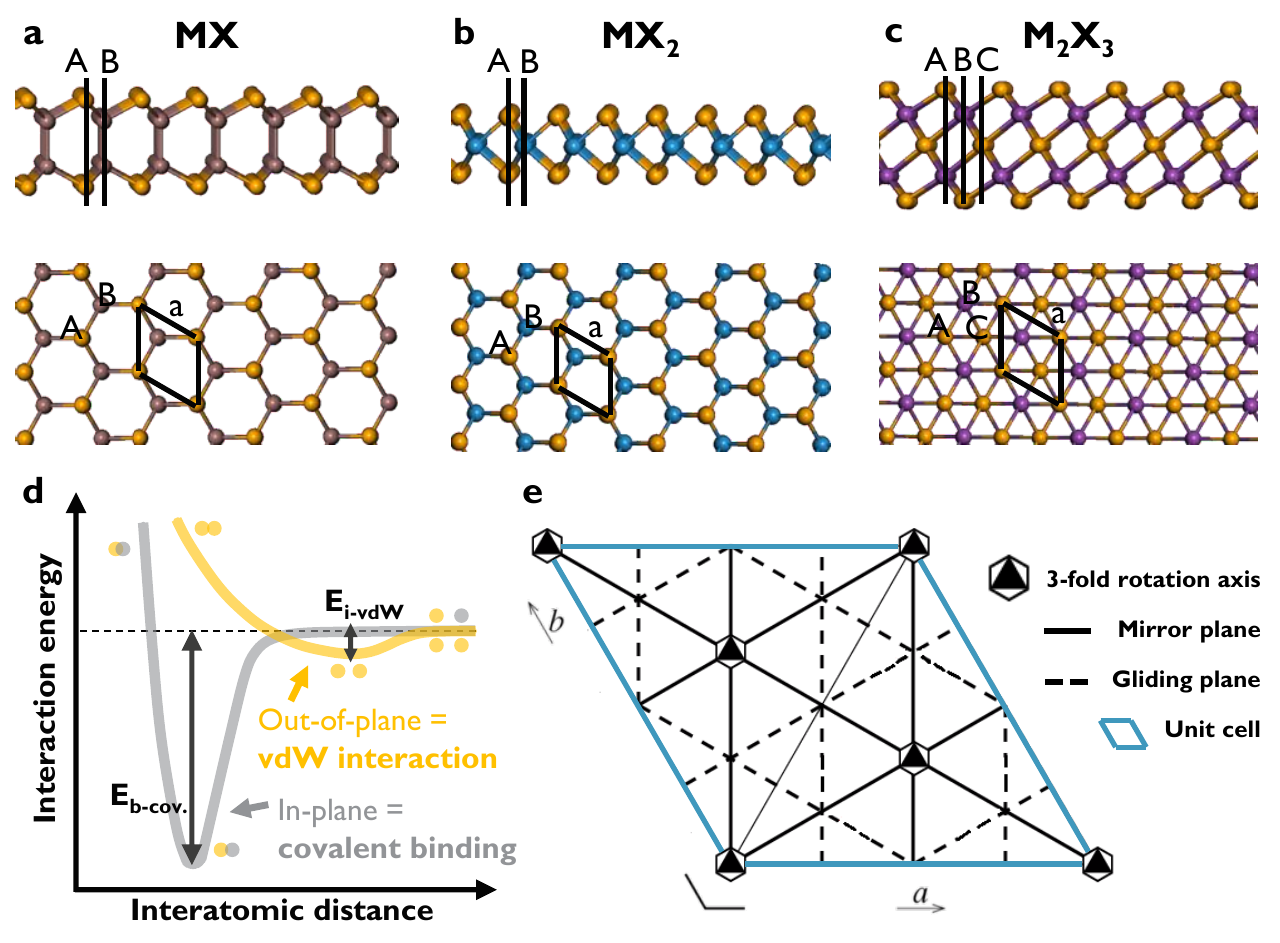}
  \caption[1]{Crystal structure, bonding and in-plane symmetry of 2D chalcogenides. a-c) 2D crystal structures of the various 2D chalcogenide compounds using tilted side-view (top panel) and top-view (bottom panel) ball-and-stick schematic illustrations for (a) MX, (b) $\text{MX}_{\text{2}}$ and (c) $\text{M}_{\text{2}}$$\text{X}_{\text{3}}$. The A, B and C stacking positions and the hexagonal unit cells with in-plane parameters are indicated. d) Interaction energy in function of the interatomic distance of two atoms that bond covalently (gray) or through vdW interactions (yellow). e) Top-view representation of the in-plane symmetry of 2D chalcogenides. The unit cell and several 3-fold rotation axes and mirror and gliding planes are highlighted. Adapted from \cite{SpaceGroup1999}.}
  \label{figure:1.2}
\end{figure}

The layered crystal structures of the various 2D chalcogenides are displayed in Figures \ref{figure:1.2}a-c. The cross-sectional representations of the MX, $\text{MX}_{\text{2}}$ and $\text{M}_{\text{2}}$$\text{X}_{\text{3}}$ structures are respectively presented in the top panel of Figures \ref{figure:1.2}a-c, the top-view representations are displayed in the bottom panel. The MX crystal structure is composed of a hexagonal metal bilayer, sandwiched between two hexagonal chalcogen planes, resulting in a total of 4 hexagonal layers. The atoms of both chalcogen planes take the A positions while the atoms of both metal planes take the B positions, which results in a top-view honeycomb-like representation (Figure \ref{figure:1.2}a). Similarly, the $\text{MX}_{\text{2}}$ triple-layer structure is composed of one hexagonal transition metal plane (B positions) sandwiched between two hexagonal chalcogen planes (A positions), yielding the top-view honeycomb representation (Figure \ref{figure:1.2}b). The $\text{M}_{\text{2}}$$\text{X}_{\text{3}}$ crystal structure, however, is composed of a quintuple-layer structure with alternating hexagonal chalcogen and hexagonal metal planes, at alternating positions (ABCAB). The occupation of all three A, B and C positions results in a top-view hexagonal-like representation (Figure \ref{figure:1.2}c). 

All sandwiched structures are covalently linked together and form the stable 2D crystal structure of the respective MX, $\text{MX}_{\text{2}}$ and $\text{M}_{\text{2}}$$\text{X}_{\text{3}}$ layers. The individual layers, however, are linked together merely through vdW bonds, resulting in the formation of a vdW gap between the various 2D layers. These vdW bonds are generally one order of magnitude weaker compared to covalent bonds (Figure \ref{figure:1.2}d). These crystal structures hence have strong intralayer interactions and weak interlayer interactions. In theory, this results in a negligible chemical reactivity in the third dimension thanks to the self-passivated surfaces at the interfaces which enables ultimate semiconductor thickness scaling down to the atomic level \cite{Pospischil2016}. As will follow, the weak interlayer vdW interactions will have important implications in the epitaxial growth of these 2D chalcogenide compounds.

The in-plane symmetry of the 2D chalcogenide single-layers, that is of crucial importance for their defect-free epitaxial growth, is visually represented in Figure \ref{figure:1.2}e \cite{SpaceGroup1999}. Here, the origin of several 3-fold rotation axes and mirror and gliding planes are depicted using a top-view representation of the 2D chalcogenides’ unit cell. As defined by symmetry, each 3-fold rotation axis collides with the intersection of three mirror planes and is equivalent to a 6-fold rotoinversion axis. The 3-fold rotational in-plane symmetry of the 2D chalcogenide single-layers is of crucial importance for the study of defect formation during epitaxial growth \cite{Lin2016}. 

\section{Epitaxy}
\label{section:E}

According to the definition of epitaxy, epitaxy involves the growth of a crystalline layer on a crystalline substrate, where the crystalline layer replicates the orientation of the underlying crystalline substrate \cite{Merriam-Webster}. The main advantage of this controlled orientational replication is that it enables high-quality, single-crystalline material growth when a defect-free, single-crystalline underlying template is used. The word \textit{"epitaxy"} is borrowed from the French word \textit{"épitaxie"} and has the etymological Greek roots of \textit{"epi"} and \textit{"taxis"} meaning respectively \textit{“above”} and \textit{“order manner”}. It was used for the first time in 1928 by the French mineralogist L. Royer in his work: \textit{“Recherches expérimentales sur l'épitaxie ou orientation mutuelle de cristaux d'espèces différentes”} \cite{Royer1928}. Nowadays, epitaxy is a well-known and widespread phenomenon, carefully and systematically applied in the semiconductor industry for the single-crystalline integration of functional materials and the realization of new device applications \cite{1389}. 

Different types of epitaxy are distinguished in the literature. When the crystalline layer is of exactly the same crystalline nature as the crystalline substrate underneath, it is referred to as homoepitaxy. When the crystalline layer is of a different nature, it is called heteroepitaxy. Homoepitaxy is generally applied to enable the manipulation of the grown material’s properties by for example the introduction of dopants during the crystal growth. Heteroepitaxy is generally applied to introduce new single-crystalline materials on existing underlying single-crystalline templates. 

In this review, epitaxy is categorized based on the nature of the interactions between the crystalline layer and the crystalline substrate \cite{Yen2019}. This results in three different categories: (1) conventional epitaxy with conventional covalent interactions between the crystalline layer and substrate, (2) vdW epitaxy with vdW interactions between the crystallites, and (3) quasi-vdW epitaxy with a (still debatable) mix of covalent and vdW interactions. The three specified types of epitaxy are discussed sequentially in the next three sections, with an emphasis on vdW and quasi-vdW epitaxy relevant for this study.

\subsection{Conventional epitaxy}
\label{subsection:CE}

\begin{figure}
  \centering
  \medskip
  \includegraphics[width=1\textwidth]{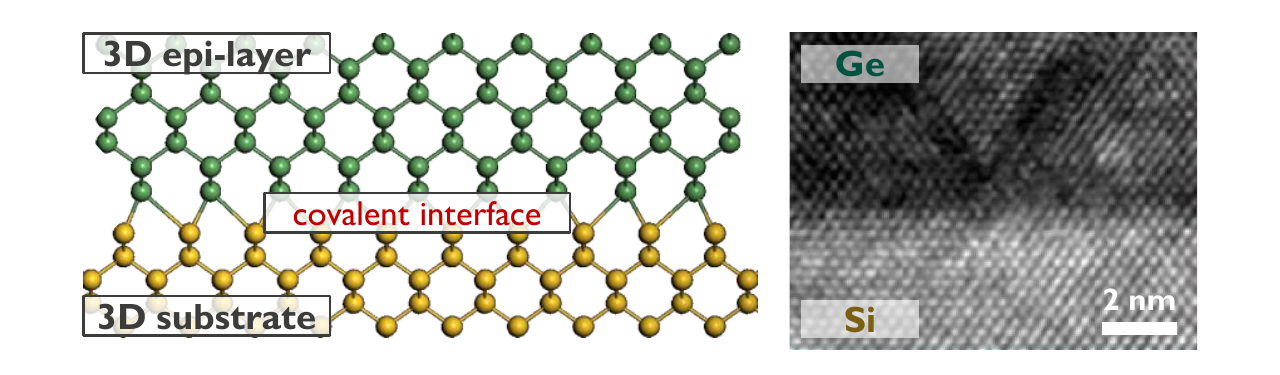}
  \caption[1]{Illustration of conventional heteroepitaxy. Left, cross-sectional ball-and-stick schematic illustration of the heterointerface of Ge grown on Si. Right, cross-sectional TEM characterization of the Ge/Si heterointerface. Adapted from \cite{Epitaxy2019}.}
  \label{figure:1.3}
\end{figure}

Until today, conventional epitaxy is one of the most critical fields in semiconductor material research. When studied and performed properly, it can enable the integration of new conventional materials that encompass new functionalities \cite{Merckling2018}. For conventional epitaxy to occur, a conventional (3D-crystal) epi-layer and a conventional (3D-crystal) epi-substrate are required. The enforced combination of both the 3D-crystals, results in a strong coupling between the epi-layer and epi-substrate owing to the covalent nature at the interface. For conventional homoepitaxy, this strong interlayer coupling generally results in a low defect density ($<$1E3 $\text{cm}^{\text{-2}}$), while for conventional heteroepitaxy the realization of such a low defect density is typically more challenging resulting from the materials' mismatch \cite{Merckling2019}. In Figure \ref{figure:1.3}, an example of conventional heteroepitaxy is highlighted for the specific epitaxial system of germanium on silicon, using both a schematic ball-and-stick illustration (left) and a cross-sectional TEM image (right) of the heterointerface \cite{Epitaxy2019}. The 3D crystal structure of both the substrate and epi-layer as well as the covalent nature of the heterointerface are clearly observed in both the ball-and-stick schematics and the TEM image. 

The major challenge of conventional heteroepitaxy is the mismatch between the epi-layer and epi-substrate material. This involves both the mismatch in lattice parameter (a) and the mismatch in coefficient of thermal expansion (CTE, $\alpha$) \cite{Merckling2018}. These mismatches generally lead to the formation of strain into the epi-layers, respectively lattice strain defined as: $\varepsilon_{\text{lattice}}$=($\text{a}_{\text{s}}$-$\text{a}_{\text{l}}$)/$\text{a}_{\text{l}}$, and thermal strain defined as: $\varepsilon_{\text{thermal}}$=$\Delta$$\alpha$$\Delta$$T$. These different strains are generally responsible for a high density of defects in the epi-layer, since defects typically result from strain relaxation (misfit dislocations) during both the heteroepitaxy and the cooldown process from the growth temperature, and can thread through the complete thickness of the epi-layer (threading dislocations) \cite{ElKazzi2012}. 

Many cases of conventional heteroepitaxy are studied in the literature such as germanium on silicon \cite{Epitaxy2019, Hu2009}, III-V materials on silicon \cite{ElKazzi2012, Ziyang2017, Sijia2015}, sapphire \cite{Hu2019} or III-V \cite{Yan2018}, oxides on oxides and on silicon \cite{Sonsteby2020, Merckling2007, Hsu2017a}, ... However, since the topic of this work involves the epitaxy of vdW materials, more attention is given to the next two categories \cite{Walsh2017b}.  

\subsection{Van der Waals epitaxy}
\label{subsection:VDWE}

\begin{figure}
  \centering
  \medskip
  \includegraphics[width=1\textwidth]{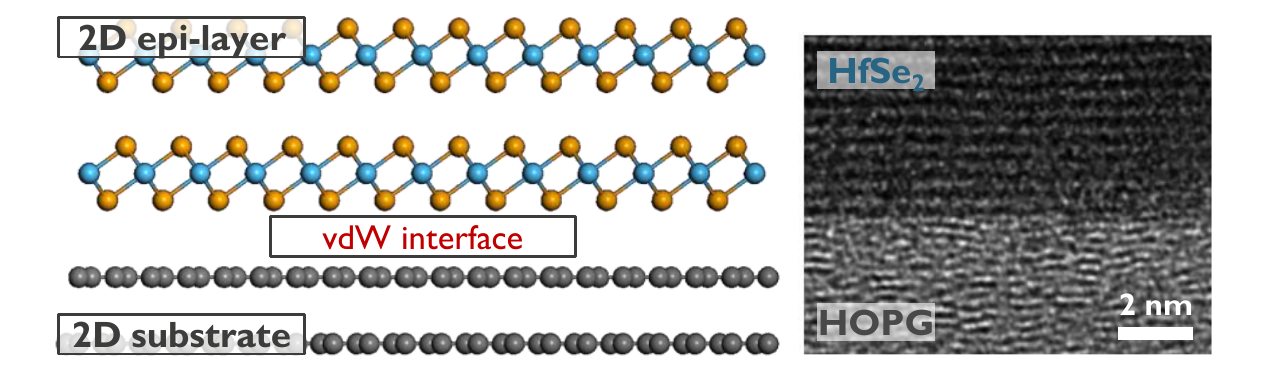}
  \caption[1]{Illustration of vdW heteroepitaxy. Left, cross-sectional ball-and-stick schematic illustration of the heterointerface of $\text{HfSe}_{\text{2}}$ grown on HOPG. Right, cross-sectional TEM characterization of the $\text{HfSe}_{\text{2}}$/HOPG heterointerface. Adapted from \cite{Yue2015a}.}
  \label{figure:1.4}
\end{figure}

Van der Waals epitaxy involves the growth of a vdW material on a vdW substrate. It was described for the first time in the 1990s by the group of A. Koma (University of Tokyo, Japan) \cite{Koma1999, Ohuchi1990}. The major difference with conventional epitaxy is that in this case, at the interface, no covalent bonding is possible due to the vdW nature of both the epi-layer and the epi-substrate. Hence, the interactions that govern the epitaxy processes are vdW interactions which are generally one order of magnitude weaker as compared to covalent interactions (see also Figure \ref{figure:1.2}d) \cite{wang2017}. This has important implications for both the homo- and heteroepitaxy processes. In Figure \ref{figure:1.4}, an example of vdW heteroepitaxy is highlighted for the specific epitaxial system of $\text{HfSe}_{\text{2}}$ on HOPG (highly oriented pyrolytic graphite), using both a schematic ball-and-stick illustration (left) and a cross-sectional TEM image (right) of the heterointerface \cite{Yue2015a}. The layered 2D structure of both the substrate and epi-layer as well as the vdW nature of the heterointerface are clearly seen in both the ball-and-stick schematics and the TEM image. The details about defect formation in vdW epitaxy will be comprehensively discussed in Section \ref{subsection:VDWE}.

\subsection{Quasi van der Waals epitaxy}
\label{subsection:QVDWE}

\begin{figure}
  \centering
  \medskip
  \includegraphics[width=1\textwidth]{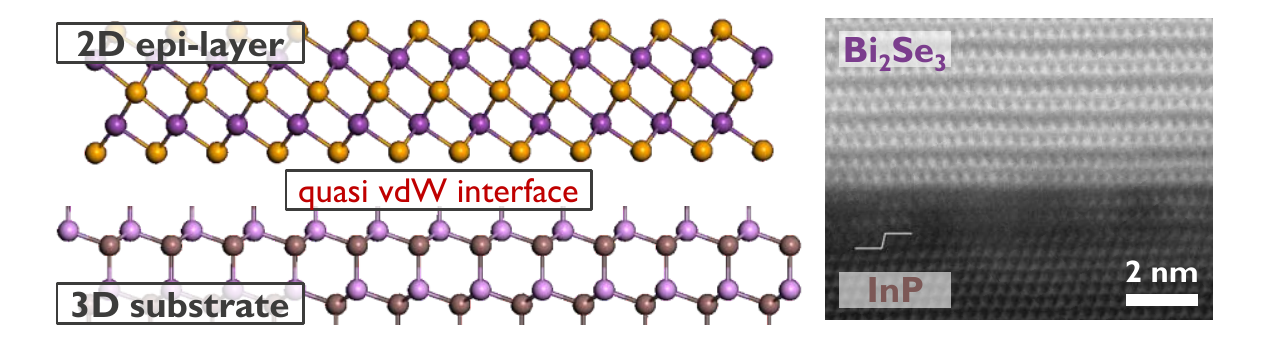}
  \caption[1]{Illustration of quasi-vdW heteroepitaxy. Left, cross-sectional ball-and-stick schematic illustration of the heterointerface of $\text{Bi}_{\text{2}}$$\text{Se}_{\text{3}}$ grown on InP. Right, cross-sectional TEM characterization of the $\text{Bi}_{\text{2}}$$\text{Se}_{\text{3}}$/InP heterointerface. Adapted from \cite{Tarakina2014}.}
  \label{figure:1.5}
\end{figure}

Similarly, as vdW epitaxy, quasi-vdW epitaxy was described and defined for the first time by the group of A. Koma \cite{Koma1999}. It involves the epitaxial growth of a vdW material on a conventional 3D crystalline substrate (or vice versa \cite{Littlejohn2017}). Quasi-vdW epitaxy is \textit{a priori} heteroepitaxy, since the very same material cannot have both a 2D and 3D crystal structure. The heterointerface in quasi-vdW heteroepitaxy is generally said to have a quasi-vdW nature, since the interactions at the heterointerface are believed not to encompass a pure vdW nor a pure covalent nature. This, however, is a topic of debate and can depend from system to system. In its essence, quasi-vdW epitaxy only involves the growth of the very first vdW ML, since subsequent MLs essentially grow through vdW homoepitaxy (and not anymore through quasi-vdW heteroepitaxy). However, this is generally overlooked (or neglected) in the literature. Therefore, in the following, we also include multilayer epitaxy in the discussion of quasi-vdW heteroepitaxy. The phenomena that are seen in these multilayer epitaxies are then associated to quasi-vdW epitaxy, and not to vdW epitaxy as described in the previous section. In Figure \ref{figure:1.5}, an example of multilayer quasi-vdW heteroepitaxy is highlighted for the specific epitaxial system of $\text{Bi}_{\text{2}}$$\text{Se}_{\text{3}}$ on InP(111)B substrate, using both a schematic ball-and-stick illustration (left) and a cross-sectional TEM image (right) of the heterointerface \cite{Tarakina2014}. The 3D crystal structure of the substrate and the 2D layered structure of the epi-layer as well as the incompatible and complex covalent/vdW nature of the heterointerface are clearly highlighted in both the ball-and-stick schematics and the TEM image. The details about defect formation in quasi-vdW epitaxy will be comprehensively discussed in Section \ref{subsection:QVDWE}.

\subsection{Overview}
\label{subsection:S}

The key aspects of the epitaxy processes that will be discussed in the following (Section \ref{section:EO2DC}) are schematically summarized in Table \ref{table:1.2}. The epitaxy processes are separated into the three categories of conventional epitaxy, vdW epitaxy and quasi-vdW epitaxy, and are divided from their homo- and heteroepitaxy processes. The focus is set on the characteristics such as in-plane registry, stacking faults and strain, that will be thoroughly discussed in the next section as regards the epitaxial growth of 2D chalcogenides.

\begin{table}
  \centering
  \medskip
  \caption[1]{Concise schematic comparison of the discussed epitaxy processes and their typical characteristics with respect to in-plane registry, stacking faults and strain.}
 \includegraphics[width=1\textwidth]{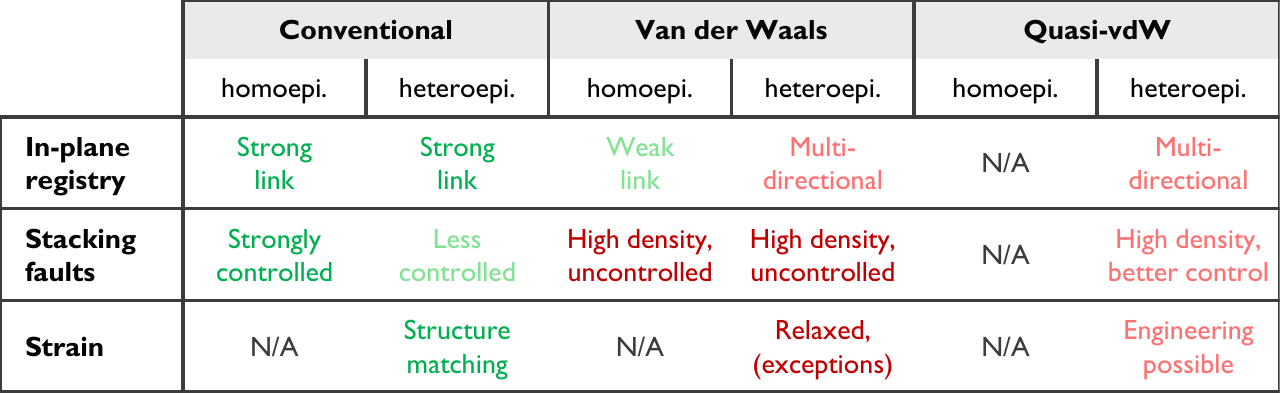} 
 \label{table:1.2}
\end{table}

The in-plane epitaxial registry of a 3D crystalline layer in conventional epitaxy (both homo- and hetero-) is generally well-controlled and governed by strong covalent linking (Table \ref{table:1.2}). This is generally more challenging in vdW epitaxy where the 2D crystalline layer is only linked through weak vdW interactions. This can give rise to multidirectional in-plane registries in vdW heteroepitaxy (Section \ref{subsubsection:ER}). The complex nature of the heterointerface in quasi-vdW heteroepitaxy similarly can give rise to multidirectional in-plane registries of the grown vdW layer (Section \ref{subsubsection:EA}).

The formation of stacking faults is generally well-controlled in conventional homoepitaxy (Table \ref{table:1.2}). The introduction of a lattice and/or structural mismatch between the 3D crystalline layer and the 3D crystalline substrate makes the control on stacking faults more challenging in conventional heteroepitaxy. The formation of stacking faults is frequently observed in vdW homo-, vdW hetero- and quasi-vdW heteroepitaxy, but is slighly better controlled in the latter (Sections \ref{subsubsection:SF} and \ref{subsubsection:SFR}).
 
Lastly, the generation of strain is generally reported in convention heteroepitaxy through strong lattice and CTE matching resulting from the covalent nature of the heterointerface (Table \ref{table:1.2}). For vdW materials, strain engineering is more challenging resulting from the weaker interactions at the heterointerfaces. In vdW heteroepitaxy, strain generation through lattice matching is researched and generally reported to be relaxed (Section \ref{subsubsection:S}). In quasi-vdW heteroepitaxy, the probing and observation of strain directly at the quasi-vdW heterointerface remains challenging but is reported in several cases, making strain engineering in 2D/3D heterostructures possible (Section \ref{subsubsection:SE}). 

\section{Epitaxial growth of 2D chalcogenides}
\label{section:EO2DC}

\subsection{Van der waals epitaxy}
\label{subsection:VDWE}

\subsubsection{Epitaxial registry}
\label{subsubsection:ER}

The orientational replication of vdW layers on vdW substrates through vdW interaction is successfully demonstrated in many different cases \cite{Tian2017}. These range from MBE grown $\text{MX}_{\text{2}}$ materials on different 2D substrates (HOPG \cite{Jiao2015a, Liu2015b, Ma2017a, Yue2017}, graphene \cite{Alvarez2018, Liu2018, Zhang2016, Quang2017, Dau2018b, LeQuang2018, Ehlen2019, Hall2018a}, $\text{MX}_{\text{2}}$ \cite{Mortelmans2020, Mortelmans2020a, W.Mortelmans2020}, \textit{h}-BN \cite{Fu2017, Loh2018}, mica \cite{Vergnaud2020}, ...), to many other 2D materials such as for example $\text{M}_{\text{2}}$$\text{X}_{\text{3}}$ \cite{Boschker2016, Chen2017, Liu2012, Li2015a, Park2016, Vermeulen2018}, \textit{h}-BN \cite{Elias2019} and graphene \cite{Summerfield2016}. Also many other different combinations using different growth techniques like for example MOVPE vdW epitaxy on graphene \cite{Azizi2015, Eichfeld2015, Rigosi2018}, CVD vdW epitaxy on graphene \cite{Kastl2018, Lin2014}, \textit{h}-BN \cite{Okada2014, Wang2015, Yan2015, Gehring2012, Liu2016} and mica \cite{Peng2012}, and ALD vdW epitaxy \cite{Hao2019}, have shown successes in demonstrating controlled, aligned growth. Interestingly, the effect of lattice mismatch and mismatch in thermal expansion coefficient is found in many cases less critical caused by the weaker interlayer coupling between the epi-layers and growth substrate \cite{Fu2017, Loh2018, Li2015a}. This enables the growth of highly lattice mismatched heterostructures, very promising for nano- and optoelectronic applications. Hence, the weaker vdW coupling results in the absence of misfit and threading dislocations, that are generally seen in conventional heteroepitaxy. In vdW heteroepitaxy, the defects that are commonly observed are 1D defects such as grain boudaries (GBs) originating from coalesced nuclei, and 0D defects such as vacancies and interstitials originating from imperfect growth conditions and intrinsic limitations \cite{Addou2015, Lin2018a, Zhao2018, NalinMehta2020, Edelberg2019, Hong2017a, VanDerZande2013}. Defect quantification of vdW epitaxy however remains extremely challenging and remains limited to few reports \cite{Mortelmans2020}. This is mainly resulting from the complex material systems combined with the finite resolution of characterization techniques, hampering accurate defect quantification. 

\begin{figure}[!t]
  \centering
  \medskip
  \includegraphics[width=1\textwidth]{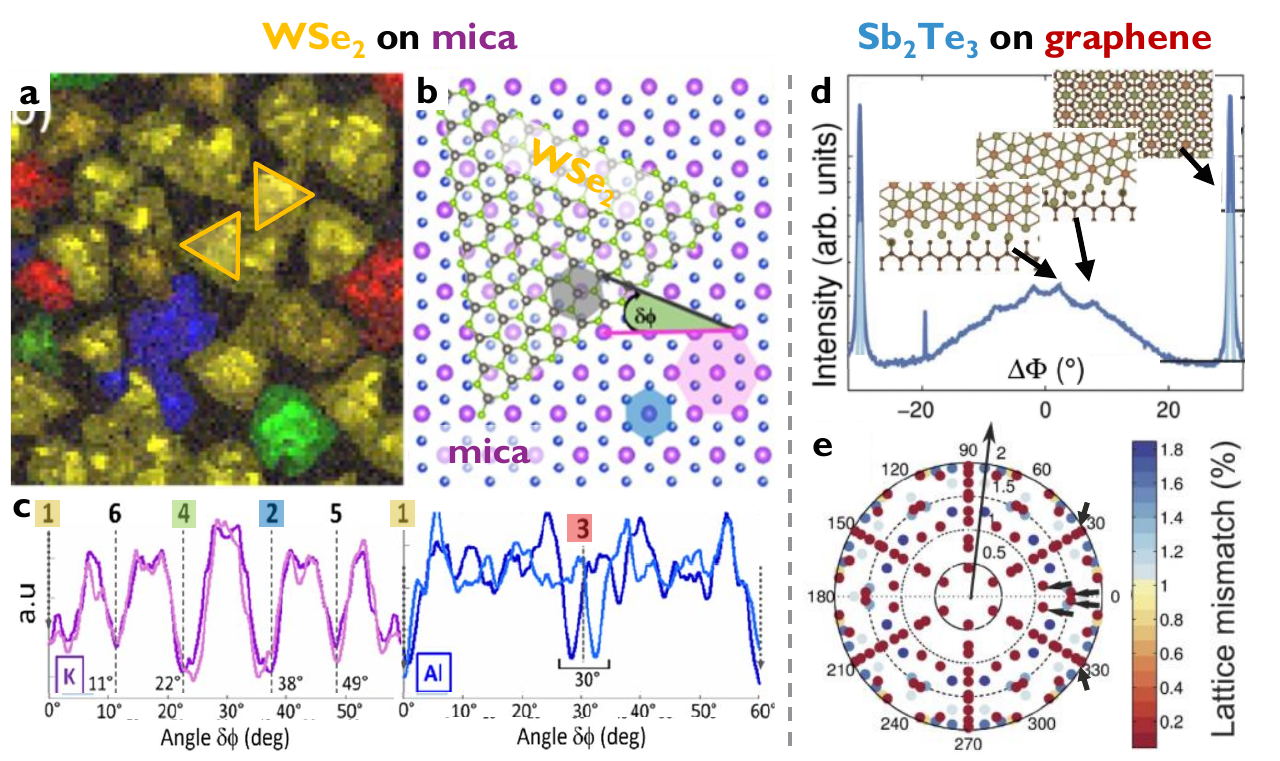}
  \caption[1]{In-plane registry of 2D chalcogenides in vdW epitaxy a-c) $\text{WSe}_{\text{2}}$ on mica. a) Color-coded plan-view TEM image where the different colors correspond to different in-plane orientations of the $\text{WSe}_{\text{2}}$. b) Top-view schematic representation of the $\text{WSe}_{\text{2}}$ on mica. The mica substrate is represented using purple atoms for K and blue atoms for Al. c) Cumulative in-plane distance between the W atoms of the $\text{WSe}_{\text{2}}$ with respect to the K atoms (left) and Al atoms (right) in function of the in-plane orientation of the $\text{WSe}_{\text{2}}$ crystal. The minima are numbered and color-coded as in agreement with the colors used in (a). d-e) $\text{Sb}_{\text{2}}$$\text{Te}_{\text{3}}$ on graphene. d) XRD $\Phi$-scan on the $\text{Sb}_{\text{2}}$$\text{Te}_{\text{3}}$(015) diffraction peak for the film grown on bilayer graphene. The relative orientations between the $\text{Sb}_{\text{2}}$$\text{Te}_{\text{3}}$ film and the graphene layer are indicated in the schematics. C, Te and Sb atoms are represented by brown, green and orange circles, respectively. For clarity reasons the crystal lattices at small misorientation angles are not overlapping. e) The absolute lattice mismatch of $\text{Sb}_{\text{2}}$$\text{Te}_{\text{3}}$ as a function of the in-plane angle and distance between the coincidence lattice sites with respect to graphene. For clarity reasons, lattice mismatches exceeding 2 \% are omitted from the graph. Adapted from \cite{Vergnaud2020, Boschker2016}.}
  \label{figure:1.6}
\end{figure}

One remarkable feature of vdW heteroepitaxy is the observation of multiple in-plane epitaxial orientations of the vdW crystals \cite{Vergnaud2020, Alvarez2018, Yan2015, Lu2017, Diaz2015}. Such a feature is for example observed for the MBE growth of $\text{WSe}_{\text{2}}$ on mica vdW substrates and presented in Figure \ref{figure:1.6} \cite{Vergnaud2020}. In Figure \ref{figure:1.6}a, a color-coded plan-view TEM image of this epitaxial structure is shown, where the different colors represent the different in-plane epitaxial orientations of the $\text{WSe}_{\text{2}}$ crystals. The observed various orientations are found to correlate with the mica surface structure (Figure \ref{figure:1.6}b), and are linked with the average in-plane distance between the W atoms of the epi-layer and the K and Al atoms of the mica surface (Figure \ref{figure:1.6}c). The in-plane orientations of the $\text{WSe}_{\text{2}}$ that have the smallest cumulative distances, are concluded to represent the preferred in-plane epitaxial registries. This hence reveals that weak vdW interlayer interactions during the vdW epitaxy process can give rise to multiple in-plane epitaxial registries, something which has also been previously reported for very specific conventional heteroepitaxial systems like crystalline oxides on silicon substrates \cite{Merckling2008, Merckling2007a}. 

In a similar case, the MBE growth of $\text{Sb}_{\text{2}}$$\text{Te}_{\text{3}}$ on graphene is highlighted, which also yields multidirectional orientations of the 2D crystals epitaxially grown on the single-crystalline 2D graphene surface \cite{Boschker2016}. This is measured by polar XRD analyses where the multiple diffraction peaks correspond to the multiple in-plane orientations of the 2D crystals (Figure \ref{figure:1.6}d). The variously observed in-plane orientations are, similarly as for the case of $\text{WSe}_{\text{2}}$ on mica, explained by the better commensurate lattice formation of the 2D $\text{Sb}_{\text{2}}$$\text{Te}_{\text{3}}$ crystals on the 2D graphene surface (Figure \ref{figure:1.6}e). Hence, these two examples demonstrate that commensurate lattice matching in vdW heteroepitaxy can play an important role in the in-plane epitaxial alignment and registry of the 2D vdW crystals. 

\subsubsection{Stacking faults}
\label{subsubsection:SF}

Besides multiple in-plane registries, the formation of stacking faults such as 60$\degree$ twins in vdW epitaxy is a major but generally hidden concern, and seems invariant from the applied growth technique and/or the vdW substrate \cite{Vergnaud2020, Vishwanath2015a, Diaz2016a, Lee2013, Ma2017a, Yue2017, Hall2018a, Chen2017, Azizi2015, Eichfeld2015, Kastl2018, Lin2014, Okada2014, Wang2015, Yan2015, Gehring2012, Mortelmans2020, Mortelmans2020a, W.Mortelmans2020}. For example, such a phenomenon is similarly observed in Figure \ref{figure:1.6}a, where the yellow color actually represents a binary in-plane registry of the $\text{WSe}_{\text{2}}$ crystals, with a 60$\degree$ relative in-plane rotation with respect to each other (overlaid yellow triangles). 

\begin{figure}[!t]
  \centering
  \medskip
  \includegraphics[width=1\textwidth]{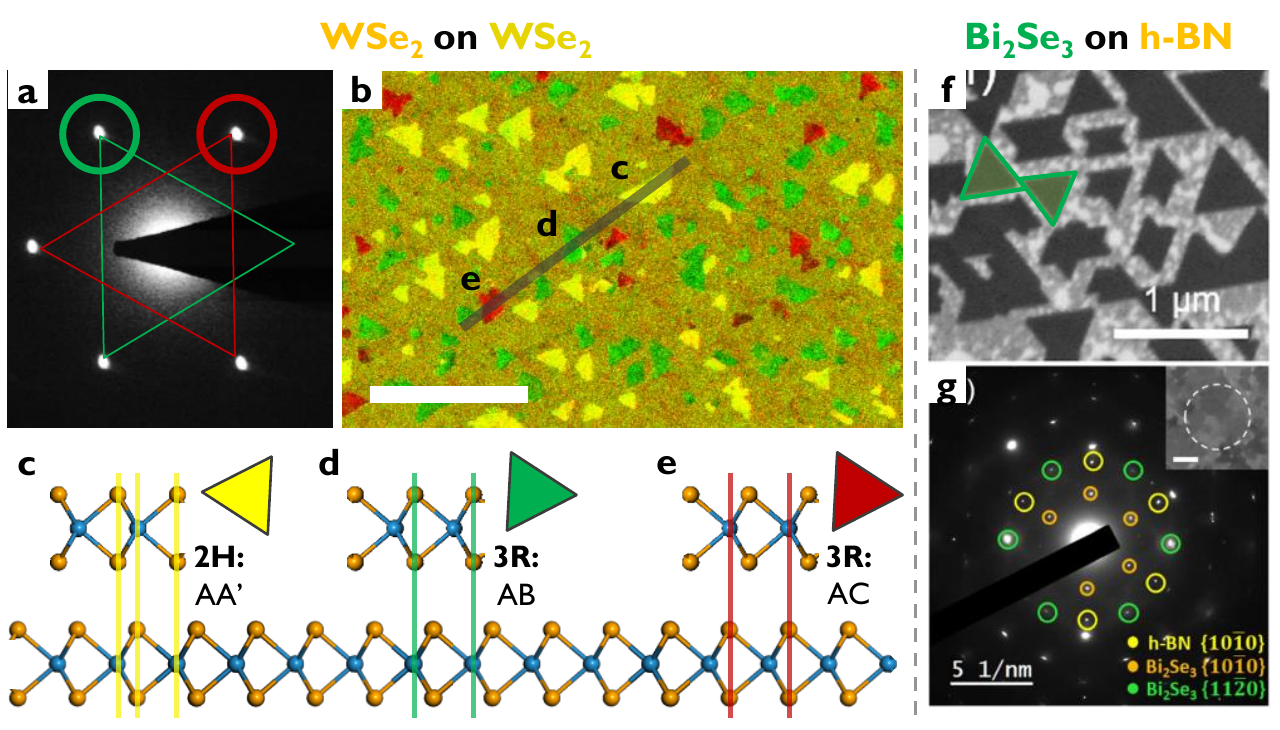}
  \caption[1]{Stacking faults of 2D chalcogenides in vdW epitaxy a-e) $\text{WSe}_{\text{2}}$ on $\text{WSe}_{\text{2}}$. a-b) TEM characterization of $\text{WSe}_{\text{2}}$ vdW homoepitaxy. a) SAEDP where the green and red triangles assemble the diffractions spots that correspond to the different \{1$\bar{1}$00\} families. The encircled green and red diffraction spots are used for the reconstruction of the image presented in (b). b) Color-coded DF-TEM image of the $\text{WSe}_{\text{2}}$ homoepitaxy. The different colors represent the various bilayer stacking configurations. The yellow color corresponds to the 2H:AA' configuration. The green and red colors correspond to respectively the 3R:AB and 3R:AC bilayer stacking configuration. Scale bar is 100 nm. c-e) Cross-sectional schematic representations of respectively the 2H:AA', 3R:AB and 3R:AC bilayer stacking configurations. f-g) $\text{Bi}_{\text{2}}$$\text{Se}_{\text{3}}$ on \textit{h}-BN. f) SEM characterization of the vdW heteroepitaxy highlighting the twinned growth (overlaid green triangles). g) TEM characterization of the $\text{Bi}_{\text{2}}$$\text{Se}_{\text{3}}$ thin film representing the SAEDP obtained from the area shown in the inset. Scale bar is 500 nm. Adapted from \cite{Mortelmans2020, Park2016}.}
  \label{figure:1.7}
\end{figure}

The presence of these stacking faults (60$\degree$ twins) and others, is studied in more depth for the curious case of $\text{WSe}_{\text{2}}$ MBE vdW homoepitaxy \cite{Mortelmans2020}. In Figure \ref{figure:1.7}a, the selected area electron diffraction pattern (SAEDP) of the $\text{WSe}_{\text{2}}$ nucleated and grown on an exfoliated single-crystalline $\text{WSe}_{\text{2}}$(0001) bilayer is presented, demonstrating the perfect registry between the epitaxial $\text{WSe}_{\text{2}}$ crystals and the $\text{WSe}_{\text{2}}$ flake surface (single-dot-like hexagonal diffraction pattern). However, the color-coded Dark-Field (DF-)TEM image that is sensitive to the $\text{WSe}_{\text{2}}$ bilayer stacking configuration clearly reveals three different bilayer stacking configurations: 2H:AA', 3R:AB and 3R:AC, hence the formation of stacking faults (Figure \ref{figure:1.7}b). These are respectively presented in Figures \ref{figure:1.7}c-d-e using cross-sectional ball-and-stick schematics in agreement with their occurrence as highlighted by the black line in Figure \ref{figure:1.7}b. The presence of these different bilayer stacking configurations, combined with the 3-fold rotational symmetry of the $\text{WSe}_{\text{2}}$ (and 2D chalcogenides in general), systematically results in highly defective vdW (homo-)epitaxial growth upon crystal coalescence \cite{Mortelmans2020, Grundmann2011}. Moreover, in this work \cite{Mortelmans2020}, the density of these defects is quantified for the first time using a universal nucleation and growth model for vdW epitaxy that relies on the nucleation density and the control on the bilayer stacking, and yielded numerical defect density values for closed MLs as large as $\sim$1E10 $\text{cm}^{\text{-2}}$ for the case of $\text{WSe}_{\text{2}}$ vdW homoepitaxy \cite{Mortelmans2020}.
 
Similarly, also for heteroepitaxial cases like $\text{Bi}_{\text{2}}$$\text{Se}_{\text{3}}$ MBE vdW epitaxy on \textit{h}-BN surface, 60$\degree$ twins are systematically observed in the nucleation studies \cite{Park2016}. This is presented in Figures \ref{figure:1.7}f-g using plan-view TEM analyses. Hence, it is concluded that these phenomena are generally observed in the vdW epitaxy of various 2D materials on various 2D surfaces, and to date no strategies to overcome the formation of these stacking faults in pure vdW epitaxy are being reported so far, apart from a specific case of $\text{MoS}_{\text{2}}$ MBE on \textit{h}-BN \cite{Fu2017}. 

\subsubsection{Strain}
\label{subsubsection:S}

\begin{figure}[!t]
  \centering
  \medskip
  \includegraphics[width=1\textwidth]{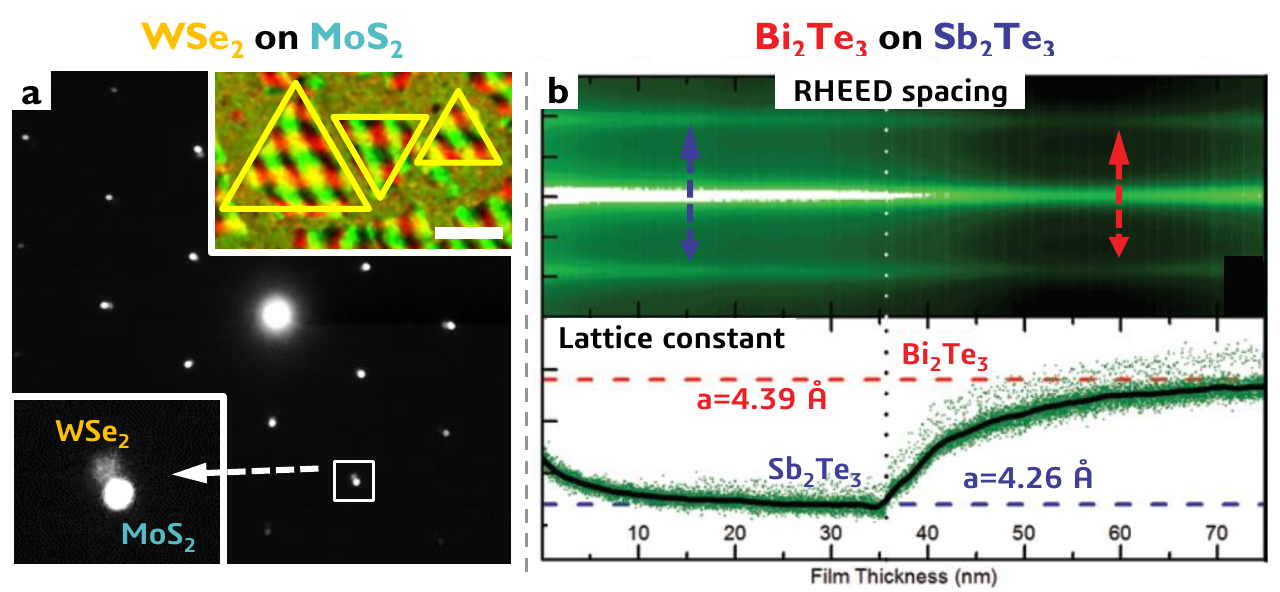}
  \caption[1]{Strain through lattice matching of 2D chalcogenides in vdW epitaxy a) $\text{WSe}_{\text{2}}$ on $\text{MoS}_{\text{2}}$. PV-TEM characterization of the $\text{WSe}_{\text{2}}$ vdW heteroepitaxy showing the SAEDP with at the bottom inset the zoom of the pattern revealing the presence of two discreate diffraction spots and at the top inset the color-coded DF-TEM image showing the moiré pattern. Both insets emphasize the absence of strain in the vdW heterostructure. Scale bar in the top inset is 20 nm. b) $\text{Bi}_{\text{2}}$$\text{Te}_{\text{3}}$ on $\text{Sb}_{\text{2}}$$\text{Te}_{\text{3}}$. Top panel) RHEED characterization of the bilayer $\text{Bi}_{\text{2}}$$\text{Te}_{\text{3}}$-$\text{Sb}_{\text{2}}$$\text{Te}_{\text{3}}$ film deposition showing the measured peak spacing that is inversely related to the lattice parameter with an accuracy of $\pm$ 0.01 \AA. Bottom panel) Presentation of the extracted measured lattice spacing as well as the bulk values (dashed lines) demonstrating that when the deposition starts (vertical dotted line), the film clearly doesn’t relax immediately to the bulk value which proves the film is grown initially in a strained state. Adapted from \cite{Mortelmans2020a, Vermeulen2018}.}
  \label{figure:1.8}
\end{figure}

In vdW epitaxy, generally, the strain is fully relaxed resulting from the weak interlayer vdW coupling at the heterointerface. This is highlighted in Figure \ref{figure:1.8}a for the specific case of $\text{WSe}_{\text{2}}$ MBE nucleation on $\text{MoS}_{\text{2}}$ \cite{Mortelmans2020a}. The SAEDP of this bilayer heterostack, reveals the presence of a double-dot-like hexagonal diffraction pattern (see the zoom in bottom left inset), which confirms that the vdW heteroepitaxy is indeed relaxed and hence that the $\text{WSe}_{\text{2}}$ grows directly with its own lattice parameter (3.288 \AA) on top of the $\text{MoS}_{\text{2}}$(0001) surface with a lattice parameter of 3.160 \AA. This is further confirmed by the color-coded DF-TEM image (top right inset in Figure \ref{figure:1.8}a) where it is clearly noticed that the color within individual $\text{WSe}_{\text{2}}$ crystals changes, corresponding to altering stacking sequences within the individual crystals as a result of the relaxed growth with lattice mismatch of 4.05 $\%$ between the $\text{WSe}_{\text{2}}$ and $\text{MoS}_{\text{2}}$. Consequently, well-defined moiré patterns are observed at the bilayer regions from which the periodicity is used to successfully estimate the mismatch between both lattices. The presented analyses hence confirm that the nucleated 2D crystals are fully relaxed on the vdW surfaces, implying the absence of strain though lattice matching in vdW epitaxy.

However, in some cases, strain through lattice matching at the vdW heterointerface is reported or predicted \cite{Vermeulen2018, Summerfield2016, Woods2014, VanWijk2014, San-Jose2014}. One of these specific cases is highlighted in Figure \ref{figure:1.8}b based on $\text{Bi}_{\text{2}}$$\text{Te}_{\text{3}}$ (a = 4.39 \AA) MBE vdW heteroepitaxy on $\text{Sb}_{\text{2}}$$\text{Te}_{\text{3}}$ (a = 4.26 \AA). Here, RHEED analyses are presented that demonstrate the extended, indistinct change in lattice parameter of the grown $\text{Bi}_{\text{2}}$$\text{Se}_{\text{3}}$, confirming the strain of $\sim$3.0 \% through lattice matching at the specific location of the vdW heterointerface ($\text{Bi}_{\text{2}}$$\text{Te}_{\text{3}}$/$\text{Sb}_{\text{2}}$$\text{Te}_{\text{3}}$). Hence, strain through lattice matching via weak vdW interactions in pure vdW heteroepitaxy can therefore be established, but remains limited to few reports only.

\subsection{Quasi van der waals epitaxy}
\label{subsection:QVDWE}

\subsubsection{Epitaxial alignment}
\label{subsubsection:EA}

Similarly as for vdW epitaxy on vdW surfaces, in-plane epitaxial alignment of vdW layers on 3D crystalline substrates is demonstrated in numerous cases. The quasi-vdW heteroepitaxies performed using MBE focus on a wide-ranging set of substrates such as sapphire \cite{Mortelmans2019, Nakano2017, Dau2017, Wang2018a, Kashiwabara2019, Roy2016a, Lee2017, Levy2018}, GaAs \cite{Ohtake2017, Chen2018a, Rumaner1998, Onomitsu2016, Ohtake2020, Vishwanath2016}, AlN \cite{Aminalragia-Giamini2017, Tsoutsou2016, Xeno2015}, InP \cite{Guo2013, Tarakina2014}, Si \cite{Ohtake2019, Reqqass1996, Boschker2014, Zallo2017a, Li2013, Klein2014, Kampmeier2015}, and $\text{(Ca,Ba)F}_{\text{2}}$ \cite{Vishwanath2018, Vishwanath2015a, Fornari2016, Kriegner2017a, Bonell2017}. Other growth techniques, however, are generally less exploratory and mainly research the quasi-vdW heteroepitaxy of vdW materials on sapphire (MOVPE \cite{W.Mortelmans2020, Mo2020, Lin2018, Zhang2018, Eichfeld2016, Chiappe2018, Marx2018}, CVD \cite{Huang2014a, Chen2015, Dumcenco2015, Lan2018, Ji2015, Han2019, Zhang2019, Suenaga2018} and ALD \cite{Groven2018, Liu2017a, Hamalainen2018}). In-plane epitaxial alignment is often achieved after proper substrate preparation \cite{Mortelmans2019, Nakano2017, Wang2018a, Kashiwabara2019, Ohtake2017, Ohtake2020, Ohtake2019, Reqqass1996, Boschker2014, Zallo2017a, Dumcenco2015}. This is performed in order to engineer the quasi-vdW heterointerface as to better control the epitaxial nucleation and growth. Such preparation methods generally involve the termination of the substrate’s surface dangling bonds through surface passivation, surface reconstructions, or buffer layer growth. Some specific respective examples are the growth of $\text{MoSe}_{\text{2}}$ on Se-passivated GaAs(111)B \cite{Ohtake2017, Ohtake2020}, $\text{Ge}_{\text{1}}$$\text{Sb}_{\text{2}}$$\text{Te}_{\text{4}}$ on ($\sqrt{3}$x$\sqrt{3}$)R30$\degree$-Sb reconstructed Si(111) \cite{Zallo2017a}, and $\text{WSe}_{\text{2}}$ on Se-buffered $\alpha$-Al$_2$O$_3$(0001) substrates \cite{Nakano2017, Wang2018a, Kashiwabara2019}. However, some cases exist where epitaxial registry occurs without the careful execution of surface preparation steps. This underlines the complex and very case-specific nature of quasi-vdW heteroepitaxy, highly dependent on the grown vdW material, the used 3D substrate, the applied growth technique and the performed growth conditions. The impact of the lattice mismatch is generally agreed as negligible, but defect quantization remains challenging and underreported. 

\begin{figure}[!t]
  \centering
  \medskip
  \includegraphics[width=1\textwidth]{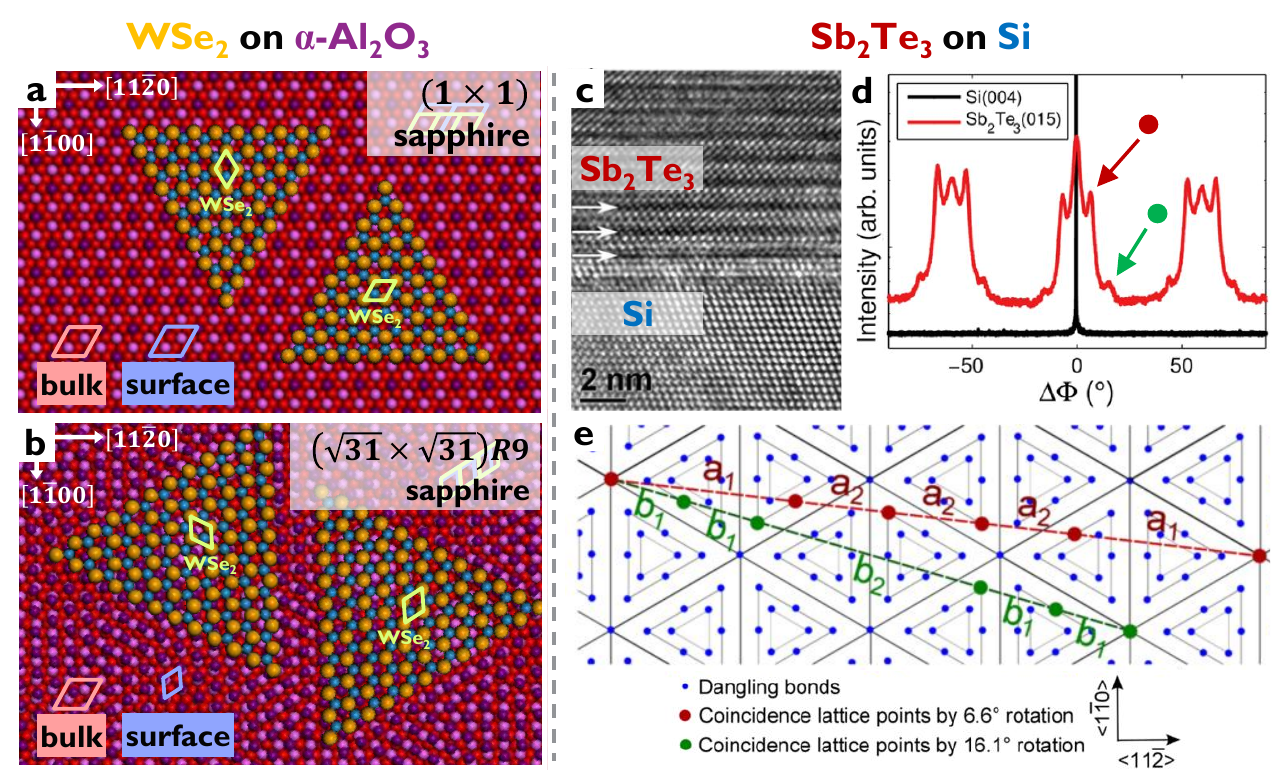}
  \caption[1]{In-plane registry of 2D chalcogenides in quasi-vdW epitaxy. a-b) $\text{WSe}_{\text{2}}$ on $\alpha$-$\text{Al}_{\text{2}}$$\text{O}_{\text{3}}$. Schematic representation of the $\text{WSe}_{\text{2}}$ quasi-vdW heteroepitaxy on (a) (1x1) and (b) ($\sqrt{31}$x$\sqrt{31}$)R9 reconstructed sapphire surfaces. The sapphire surfaces are represented by bulk oxygen (red), bulk aluminum (light purple) and surface aluminum (dark purple) atoms. The sapphire symmetry for both bulk (red diamond) and surface (blue diamond) are indicated in the bottom left corner. The symmetry of the 2D $\text{WSe}_{\text{2}}$ crystals (yellow diamonds) is indicated on the nucleated crystals itself. c-e) $\text{Sb}_{\text{2}}$$\text{Te}_{\text{3}}$ on (7x7) Si(111). c) High-resolution TEM image of the $\text{Sb}_{\text{2}}$$\text{Te}_{\text{3}}$ film grown on Si(111)-(7x7). d) $\Phi$-scans around the Si(004) and $\text{Sb}_{\text{2}}$$\text{Te}_{\text{3}}$(015) diffraction peaks for $\text{Sb}_{\text{2}}$$\text{Te}_{\text{3}}$ grown on Si(111)-(7x7). e) Schematic of the Si(111)-(7x7) surface. Each large triangle corresponds to one (un)faulted half of the reconstructed surface. The dangling bonds on the surface are marked by small blue dots. The red and green dashed lines are rotated by 6.6$\degree$ and 16.1$\degree$ with respect to the $\langle$11$\bar{2}$$\rangle$ direction respectively. The dangling bonds lying on these lines are marked with large red and green dots. The distances between these dots are indicated. Adapted from \cite{Mortelmans2019, Boschker2014}.}
  \label{figure:1.9}
\end{figure}

In quasi-vdW epitaxy, the surfaces of the (engineered) substrates are found to play a crucial role in the quasi-vdW heteroepitaxial registry of the 2D crystals. This is for example emphasized in the case of $\text{WSe}_{\text{2}}$ nucleated on (1x1) and ($\sqrt{31}$x$\sqrt{31}$)R9 reconstructed sapphire surfaces \cite{Mortelmans2019}. In Figures \ref{figure:1.9}a-b, both these surfaces are modeled including the $\text{WSe}_{\text{2}}$ nucleation, and displayed using ball-and-stick schematic representations. The unit cells of the sapphire surfaces and the 2D $\text{WSe}_{\text{2}}$ crystals are highlighted. In both cases, the in-plane epitaxial alignment is directly governed by the uppermost surface resulting in the epitaxial relation of [11$\bar{2}$0]$\text{WSe}_{\text{2}}$(0001)//[11$\bar{2}$0]$\alpha$-$\text{Al}_{\text{2}}$$\text{O}_{\text{3}}$(0001) on the (1x1) constructed sapphire surface, and interestingly an azimuthal rotation of 30$\degree$ to [1$\bar{1}$00]$\text{WSe}_{\text{2}}$(0001)//[11$\bar{2}$0]$\alpha$-$\text{Al}_{\text{2}}$$\text{O}_{\text{3}}$(0001) on the ($\sqrt{31}$x$\sqrt{31}$)R9 surface explained by the $\sim$30$\degree$ rotation of the reconstructed surface's superstructure with respect to the bulk sapphire crystals \cite{Mortelmans2019}. Coincident lattice formation controlled by the uppermost layer of surface atoms is hence largely responsible for the specific in-plane epitaxial registry. This, clearly demonstrates the crucial impact of the starting surface's symmetry and orientation on the in-plane registry of 2D vdW crystals in quasi-vdW heteroepitaxy.

Moreover, the complexity of the quasi-vdW heterointerface is clearly highlighted for the case of $\text{Sb}_{\text{2}}$$\text{Te}_{\text{3}}$ grown on (7x7) reconstructed (111)-oriented silicon substrates (Figures \ref{figure:1.9}c-e) \cite{Boschker2014}. This quasi-vdW epitaxial system is found to generate rotational domains, whose presence is explained by the formation of coincident lattices with the underlying dangling bonds of the Si(111) substrate's surface in order to reduce the lattice mismatch. In Figure \ref{figure:1.9}c, the cross-sectional TEM image of such a system is presented where the dark band at the heterointerface indeed suggests the existence of the (quasi-)vdW gap nature. Furthermore, the azimuthal X-ray diffraction (XRD) pattern of the $\text{Sb}_{\text{2}}$$\text{Te}_{\text{3}}$(015) planes clearly highlights the presence of the rotational domains since multiple diffraction peaks are observed next to the main diffraction peaks (Figure \ref{figure:1.9}d). The positions of these additional diffraction peaks are shifted by 6.6$\degree$ and 16.1$\degree$ (red and green respectively), and are directly linked with the dangling bond sub-structure of the underlying reconstructed Si(111) surface. This is demonstrated in Figure \ref{figure:1.9}e, where the rotated coincident lattices are nicely observed to match integer numbers of $\text{Sb}_{\text{2}}$$\text{Te}_{\text{3}}$ unit cells with the interspaces of the (7x7) reconstructed silicon’s surface dangling bonds ($\text{a}_{\text{1}}$/$\text{a}_{\text{2}}$, and $\text{b}_{\text{1}}$/$\text{b}_{\text{2}}$). Both the geometry and the dangling bond structure of the reconstructed surface of the 3D-crystal substrate can hence control multiple in-plane epitaxial registries of the vdW layer, which is a remarkable feature of quasi-vdW epitaxy of 2D chalcogenides \cite{Boschker2014, Ohtake2020, Eichfeld2016, Ji2015, Han2019, Liu2014a}.

\subsubsection{Stacking fault reduction}
\label{subsubsection:SFR}

The formation of stacking faults such as 60$\degree$ twins is also frequently observed in quasi-vdW epitaxy \cite{Mortelmans2019, Mo2020, W.Mortelmans2020, Boschker2014, Lee2017, Chen2018a, Ohtake2020, Vishwanath2015a, Xu2017, Lin2018, Zhang2018, Eichfeld2016, Chiappe2018, Marx2018, Dumcenco2015, Han2019, Suenaga2018, Kriegner2017a, Guo2013}. This could already be observed in Figure \ref{figure:1.9}, where the 6-fold registry and the 6-fold periodicity of the XRD pattern is inconsistent with the 3-fold in-plane symmetry of the grown 2D chalcogenides and substrates. 

\begin{figure}[!t]
\centering
\medskip
\includegraphics[width=1\textwidth]{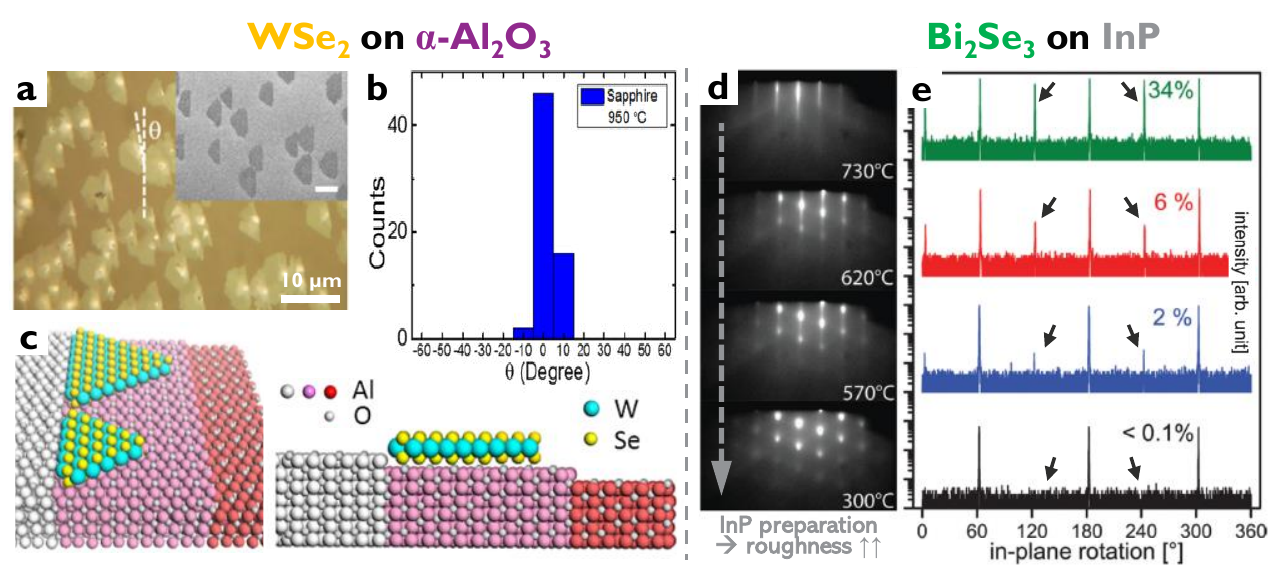}
\caption[1]{Twinning reduction of 2D chalcogenides in quasi-vdW epitaxy. a-c) $\text{WSe}_{\text{2}}$ on $\alpha$-$\text{Al}_{\text{2}}$$\text{O}_{\text{3}}$. a) AFM and OM (inset) images of the $\text{WSe}_{\text{2}}$ quasi-vdW heteroepitaxy on sapphire. b) Histogram of the orientation distribution based on the image in (a). c) Atomic models illustrating the aligned nucleation of $\text{WSe}_{\text{2}}$ on the surface steps of the c-plane sapphire substrates. d-e) $\text{Bi}_{\text{2}}$$\text{Se}_{\text{3}}$ on InP. d) RHEED patterns recorded from the InP substrate for different annealing temperatures. e) Pole scans of the $\text{Bi}_{\text{2}}$$\text{Se}_{\text{3}}$\{015\} reflections for different annealing temperatures of the InP(111)B substrates; the indicated twin volume of the crystal is determined by the peak area of the twin triplet compared to the total measured peak area. The InP\{002\} reflections (not shown) occur at the same in-plane angles as the non-suppressed peak triplet. A logarithmic scale is used in the plots. Adapted from \cite{Chen2015, Tarakina2014}.}
\label{figure:1.10}
\end{figure}

Interestingly, one particular case of twin reduction in $\text{MX}_{\text{2}}$ quasi-vdW heteroepitaxy is reported and relies on the ordering of the vdW crystals at the surface steps of the single-crystalline substrate \cite{Chen2015}. In Figures \ref{figure:1.10}a-c, this is presented for the specific case of $\text{WSe}_{\text{2}}$ CVD on c-plane sapphire substrates. Both atomic force and optical microscope images (Figure \ref{figure:1.10}a) are analyzed to result in untwined quasi-vdW heteroepitaxy (Figure \ref{figure:1.10}b), which is explained by the strong (covalent) linking of the edges of the $\text{WSe}_{\text{2}}$ crystals with the atomic steps of the 0.2$\degree$ miscut sapphire surface (Figure \ref{figure:1.10}c). This challenging but promising approach can become a potential path for twinning reduction hence intergrain defect density reduction in $\text{MX}_{\text{2}}$ quasi-vdW heteroepitaxy.

In comparison with $\text{MX}_{\text{2}}$ quasi-vdW heteroepitaxy, twin reduction in $\text{M}_{\text{2}}$$\text{X}_{\text{3}}$ quasi-vdW heteroepitaxy is notably more frequently reported. The reduction of twinning in these quasi-vdW heteroepitaxial systems generally relies on optimized growth conditions \cite{Kampmeier2015, Bonell2017}, buffer layer growth \cite{Levy2018} or the intentional introduction of surface roughness \cite{Tarakina2014}. The latter, is presented in Figures \ref{figure:1.10}d-e for the case of $\text{Bi}_{\text{2}}$$\text{Se}_{\text{3}}$ on InP(111)B, where the increased substrate’s surface roughness is probed and confirmed by \textit{in situ} RHEED characterization, and the related decreased twin formation by polar XRD analyses. In a similar way as for the case above, the twin reduction is explained by the strong linking of the 2D crystals with a 3D aspect of the growth surface (surface roughness here instead of step edges above). Hence, the introduction of 3D features in the growth surfaces seems to enable a reduction in the formation of stacking faults in quasi-vdW heteroepitaxy.

Remarkably, these various strategies for twin reduction in quasi-vdW heteroepitaxy are inexistent in the cases of vdW epitaxy. This, can potentially result from the weaker interlayer coupling in pure vdW epitaxy and/or the more challenging engineering possibilities on purely vdW templates compared to 3D crystalline surfaces for efficient twin reduction.

\subsubsection{Strain engineering}
\label{subsubsection:SE}

Like in the scenario of vdW epitaxy, quasi-vdW epitaxy is generally assumed to be fully relaxed enabling the growth of largely lattice mismatch heterostructures. This statement, however, is difficult to validate, since many reports include vdW homoepitaxy into quasi-vdW heteroepitaxy which impact the analysis and the observation of strain at the quasi-vdW heterointerface. Only few cases report on the formation of strain in the very first vdW layer in quasi-vdW heteroepitaxy \cite{Mortelmans2019, Zallo2017a, Wang2018, Sun2017, Klein2014}. 

\begin{figure}[!t]
  \centering
  \medskip
  \includegraphics[width=1\textwidth]{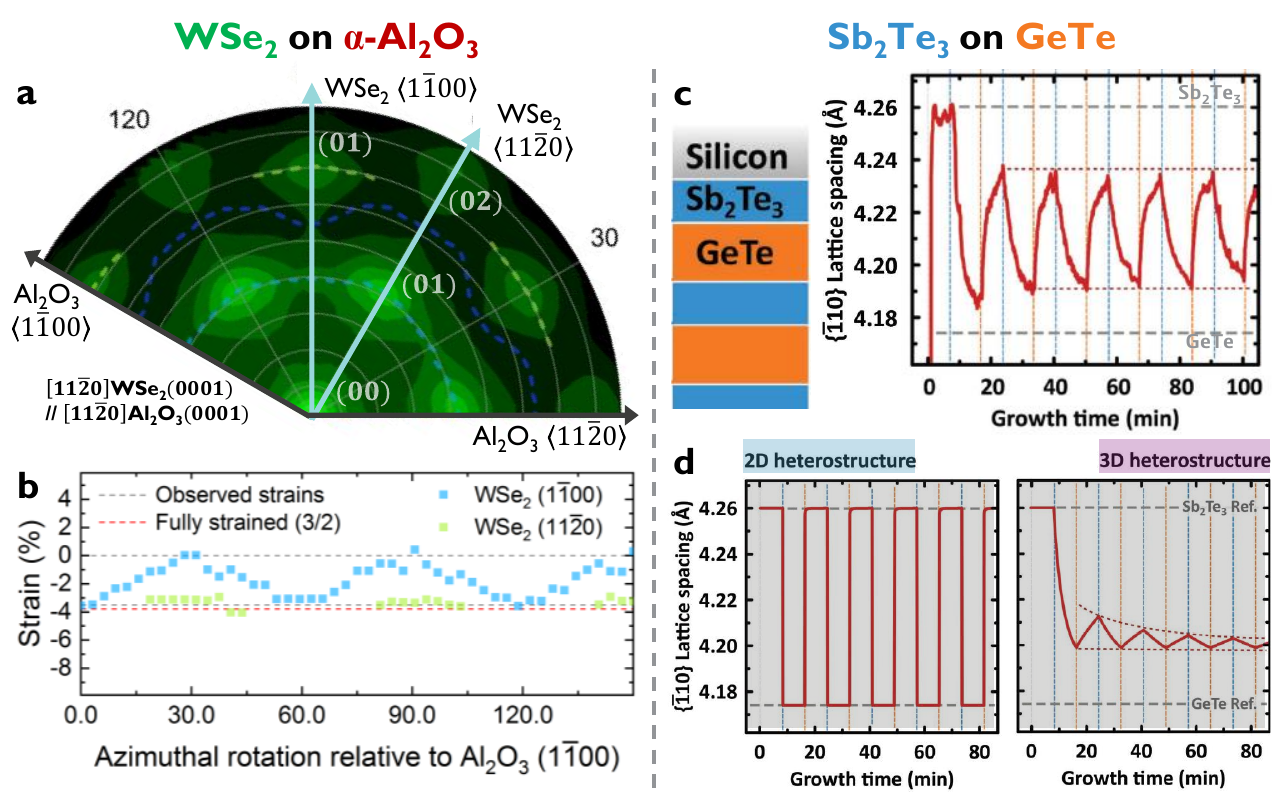}
  \caption[1]{Strain engineering of 2D chalcogenides in quasi-vdW epitaxy. a-b) $\text{WSe}_{\text{2}}$ on $\alpha$-$\text{Al}_{\text{2}}$$\text{O}_{\text{3}}$. a) Polar \textit{in situ} RHEED plot of 1ML $\text{WSe}_{\text{2}}$ film grown on sapphire. The dashed lines indicate the RHEED intensity maxima of the $\text{WSe}_{\text{2}}$ (1$\bar{1}$00) planes (light blue), the RHEED background that is extracted from the minimal RHEED intensity (dark blue), and the intensity maxima of the $\text{WSe}_{\text{2}}$ (11$\bar{2}$0) planes (green). The light blue arrows indicate the in-plane alignment of the grown 2D $\text{WSe}_{\text{2}}$ crystals relative to the bulk $\text{Al}_{\text{2}}$$\text{O}_{\text{3}}$ crystals. b) Analysis of the polar RHEED plot showing the extraction of the RHEED strain profiles in function of the azimuthal rotation relative to the $\text{Al}_{\text{2}}$$\text{O}_{\text{3}}$ (1$\bar{1}$00) planes. The strain from both the $\text{WSe}_{\text{2}}$ (1$\bar{1}$00) planes (light blue squares) and the $\text{WSe}_{\text{2}}$ (11$\bar{2}$0) planes (green squares) are plotted, and the average measured strains (gray dashed lines) and the theoretically calculated strain in case of fully lattice-matched structures with the sapphire surface (red dashed line) are presented. c-d) $\text{Sb}_{\text{2}}$$\text{Te}_{\text{3}}$ on GeTe c) Representation of the 2D/3D superlattices with the $\{$$\bar{1}$10$\}$ lattice plane spacing calculated from RHEED streak spacing monitored during the growth of 2D-$\text{Sb}_{\text{2}}$$\text{Te}_{\text{3}}$/3D-GeTe superlattices. The blue and orange dashed lines indicate the deposition of $\text{Sb}_{\text{2}}$$\text{Te}_{\text{3}}$ and GeTe respectively. d) Extreme hypothetical cases with (left) no coupling and (right) full coupling between the 2D/3D compounds. The measured in-plane lattice constant during growth displays fingerprints of both extremes, persistent oscillations (2D) and straining (3D). Adapted from \cite{Mortelmans2019, Wang2018}.}
  \label{figure:1.11}
\end{figure}

One particular case that demonstrates strain through lattice matching in single-layer quasi-vdW heteroepitaxy is highlighted in Figures \ref{figure:1.11}a-b \cite{Mortelmans2019}. Here, the MBE growth of 1 ML $\text{WSe}_{\text{2}}$ on $\alpha$-$\text{Al}_{\text{2}}$$\text{O}_{\text{3}}$(0001) is in-depth analysed using the \textit{in situ} polar RHEED characterization technique\cite{Mortelmans2019} which revealed the presence of strained $\text{WSe}_{\text{2}}$ crystals in the commensurate states of $\text{WSe}_{\text{2}}$(1$\bar{1}$00)-on-$\text{Al}_{\text{2}}$$\text{O}_{\text{3}}$(1$\bar{1}$00) (at 0$\degree$, 60$\degree$, 120$\degree$, ...), and fully relaxed crystals in the incommensurate regions (Figure \ref{figure:1.11}b). The presence of the strain in the commensurate states is explained by lattice matching of three $\text{WSe}_{\text{2}}$ units on two $\text{Al}_{\text{2}}$$\text{O}_{\text{3}}$ units, that have a reduced lattice mismatch of only -3.8 $\%$. Consequently, lattice mismatch, can, significantly impact quasi-vdW heteroepitaxy. 

Similarly, the formation of strain through lattice matching in "multilayer quasi-vdW epitaxy" is observed in MBE-grown superlattices of 2D $\text{Sb}_{\text{2}}$$\text{Te}_{\text{3}}$ and 3D GeTe compounds (Figures \ref{figure:1.11}c-d) \cite{Wang2018}. Here, the lattice parameters of the grown structures, monitored by \textit{in situ} RHEED analyses (Figure \ref{figure:1.11}c), is shown to oscillate between the boundary values of the individual relaxed layers, which is related to the non-zero coupling between the various 2D/3D compounds. This explanation is supported by simulations of completely uncoupled layers (ideal 2D heterostructure, Figure \ref{figure:1.11}d (left)) and fully covalently coupled layers (ideal 3D heterostructure, Figure \ref{figure:1.11}d (right)), where the experimental data fits in between these boundary cases. Hence, such observations open perspectives for strain engineering in 2D/3D heterostructures and in quasi-vdW heteroepitaxy in general, which is promising for future device applications where strain can play an important role in the bandgap, charge carrier mobility and topology.

\section{Generalized nucleation density benchmark}
\label{section:B}

The review presented above on the epitaxial growth of 2D chalcogenides, has learned that in both vdW and quasi-vdW epitaxy, the presence of stacking faults, mainly, combined with the general absence of lattice matching and the unintentional presence of multiple in-plane epitaxial registries, make the nucleation density a key measure for the intergrain defect density of the 2D films. Therefore, in the following, the focus is placed to a generalized nucleation density benchmark on the epitaxy of TMDs ($\text{MX}_{\text{2}}$), an important member of the 2D chalcogenide family, based on the data extracted from all reported AFM, SEM, TEM and/or STM images at the nucleation level.  

In order to enable a fair comparison from all these different data, the extracted nucleation densities are corrected for the differences in surface coverage using a simplified nucleation model as reported previously \cite{Mortelmans2020a}. This simplified nucleation model relies on the theoretically derived and experimentally confirmed correlation between the adatom diffusion coefficient and the nucleation density (Equation \ref{eq:1}) \cite{Mo1991, Oura2003a}. The correction is essential since experiments having a very different surface coverage (e.g. 20 vs 60 \%) naturally result in a very different nucleation density, dominated by the difference in the amount of grown material which would lead to an incorrect interpretation of the benchmarking study. Hence, the correction is applied according to the following equations: 

\begin{equation}
D=3\theta^2/tN^3
\label{eq:1}
\end{equation}
\begin{equation}
\theta_{30\%}=0.3/(a^2_{MX_2}\sqrt{3}/2)
\label{eq:3}
\end{equation}
\begin{equation}
D_{30\%}=D
\label{eq:4}
\end{equation}
\begin{equation}
t_{30\%}=(\theta_{30\%}/\theta)*t
\label{eq:5}
\end{equation}
\begin{equation}
N_{30\%}=\sqrt[3]{3\theta_{30\%}^2/t_{30\%}D_{30\%}}=N*\sqrt[3]{0.3/(\theta*a^2_{MX_2}\sqrt{3}/2)}
\label{eq:2}
\end{equation}

Here, D represents the diffusion constant expressed in $cm^2/s$, $\theta$ the surface coverage expressed in $\#/cm^2$, t the growth time expressed in $s$ and N the nucleation density expressed in $\#/cm^2$. The subscript 30\% represents the values at a corrected surface coverage of 30 \%, with $a_{MX_2}$ the lattice paramater of the $\text{MX}_{\text{2}}$ compound (Equation \ref{eq:3}). The value of 30 \% is chosen since most of the data are as close as possible to this value, which is a good balance between sufficient amount of crystal growth in combination with as little as possible coalesced nuclei. To do this correction, the equations assume a diffusion constant that is independent from the surface coverage (Equation \ref{eq:4}) and a linear response of the surface coverage in function of time (Equation \ref{eq:5}). This finally results in Equation \ref{eq:2}, which represents the mathematical expression of the corrected nucleation density at a surface coverage of 30 \% based on all experimental parameters like nucleation density (N), surface coverage ($\theta$) and $\text{MX}_{\text{2}}$ lattice constant ($a_{MX_2}$). 

\begin{figure}[!b]
  \centering
  \medskip
  \includegraphics[width=1\textwidth]{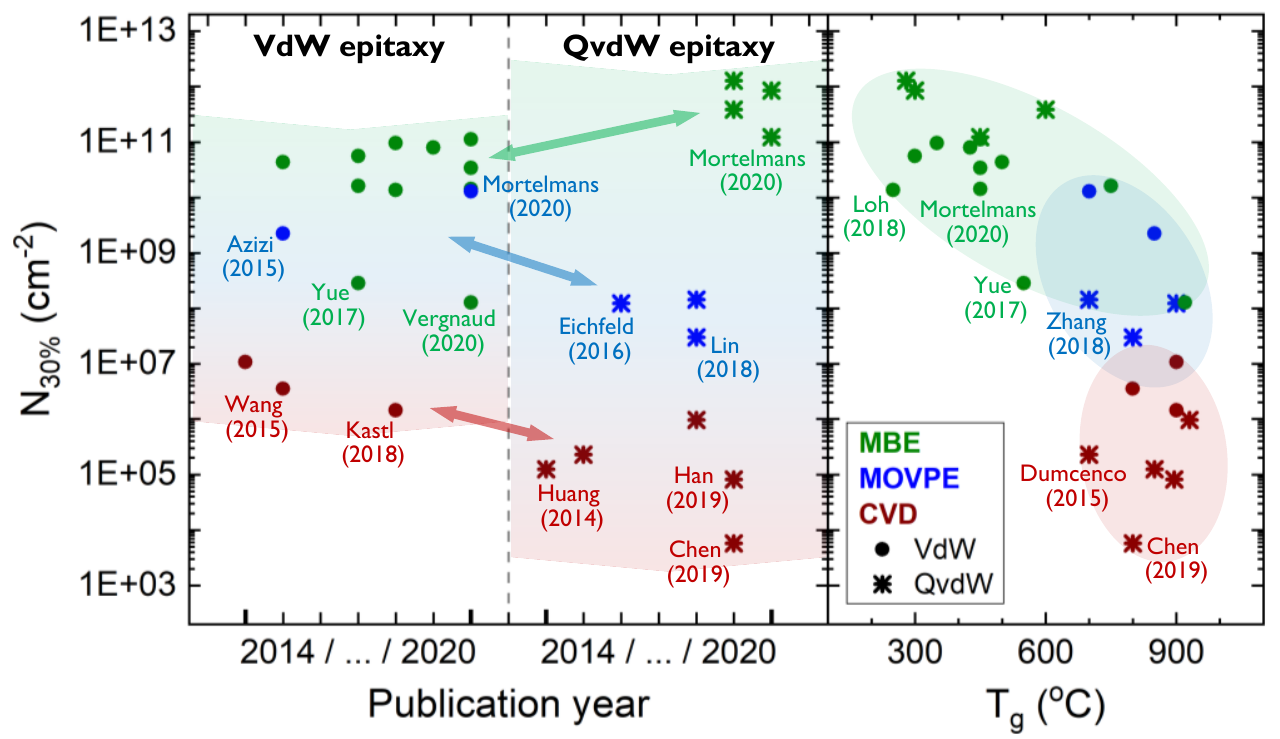}
  \caption[1]{Generalized nucleation density benchmark for the epitaxy of TMD materials. The nucleation densities are corrected to a surface coverage of 30 \% and are represented with bullets for vdW and with stars for quasi-vdW epitaxy. The green color corresponds to the MBE growth method, the blue to MOVPE and the red to CVD. The references with the lowest nucleation density within their categories are highlighted in the graph. Left panel) Nucleation density of TMD epitaxy in function of the year of publication, separated for vdW and quasi-vdW epitaxy. Right panel) Nucleation density of TMD epitaxy in function of a key growth condition that is the growth temperature.}
  \label{figure:1.12}
\end{figure}

The generalized nucleation densities at a surface coverage of 30 \% ($N_{30\%}$) for both vdW and quasi-vdW epitaxy of TMDs using the growth methods of MBE, MOVPE and CVD are presented in Figure \ref{figure:1.12}. In the left panel of the figure, the nucleation densities are presented in function of the year of publication and separated for vdW and quasi-vdW epitaxy. In the right panel of the figure, all nucleation densities are presented in function of the applied growth temperatures, ranging from 250-950 $\degree$C. 

The MBE method systematically yields the highest nucleation densities, typically above 1E10 $\text{cm}^{\text{-2}}$, for both vdW and quasi-vdW TMD epitaxy (Figure \ref{figure:1.12}, left panel). Accordingly, the closed MBE films will (generally) suffer from notably higher intergrain defect densities compared to growth methods that enable much lower nucleation densities. Interestingly, the MBE nucleation densities demonstrate systematically lower values for vdW epitaxy compared to quasi-vdW epitaxy. VdW epi-surfaces in comparison to 3D crystalline epi-surfaces are hence found advantageous in TMD MBE epitaxy. This observation can be linked with a generally enhanced surface diffusion at low energy vdW surfaces, as studied previously for various cases of TMDs\cite{Mortelmans2020a}. Two impressive outliners from Yue \textit{et al.} \cite{Yue2017} and Vergnaud \textit{et al.} \cite{Vergnaud2020} demonstrate excellent MBE vdW-growths in terms of relaxed nucleation densities in the $\sim$1E8 $\text{cm}^{\text{-2}}$ range, achieved from the exploration towards extreme MBE growth conditions like respectively ultra-low growth rates ($<$0.05 ML.$\text{h}^{\text{-1}}$) and ultra-high growth temperatures ($>$900 $\degree$C), both combined with significant chalcogen overpressures ($>$20 X/M).

However, lower nucleation densities are systematically obtained in MOVPE ($\sim$1E8-1E9 $\text{cm}^{\text{-2}}$) and CVD ($<$1E7 $\text{cm}^{\text{-2}}$) growth methods, with finest values extracted from the latter (Figure \ref{figure:1.12}, left panel). Remarkably, both chemical methods demonstrate lower nucleation density values in quasi-vdW epitaxy compared to vdW epitaxy, which is the inverse from the physical method of MBE. The inferior chemical (rather than physical) stability of vdW surfaces compared to 3D crystalline surfaces during the more demanding processes of the chemical growth methods (high temperatures ($\sim$900 $\degree$C), high pressures ($\sim$1 atm)) could explain this inverse behavior. 

Moreover, a clear link between the generalized nucleation density and the growth temperature is revealed from Figure \ref{figure:1.12} (right panel). A systematic decrease in nucleation density is observed in function of increasing growth temperature, and this generally for all applied growth techniques but with a clear separation between them. Although, other growth parameters such as growth pressure and growth rate similarly play an important role in the (quasi-)vdW epitaxies of TMD materials, this curious trend demonstrates a clear insight in the growth processes of 2D chalcogenides. 

The MBE method is shown to cover a wide range of growth temperatures positioned in the low-to-mid growth temperature window (300-600 $\degree$C for most of the reports). The observed correlation between the higher MBE growth temperature and the lower generalized nucleation density can be explained by the increased surface diffusion of adatoms at elevated growth temperatures, yielding TMD epitaxies having a lower nucleation density. Moreover, the MBE experiments that yield the lowest nucleation densities at the lowest thermal budgets (Loh \textit{et al.} \cite{Loh2018}, Mortelmans \textit{et al.} \cite{Mortelmans2020a} and Yue \textit{et al.} \cite{Yue2017}) are all of vdW epitaxy type, similarly highlighting the beneficial aspect of vdW surfaces for the MBE growth of 2D chalcogenides.

The MOVPE and CVD methods with their typically higher growth temperature windows (700-950 $\degree$C) are able to yield TMD epitaxies with nucleation densities below 1E8 and 1E4 $\text{cm}^{\text{-2}}$, respectively, which are multiple orders of magnitude lower compared to the typical nucleation densities of the MBE method ($\sim$1E11 $\text{cm}^{\text{-2}}$). Here, the best performers with lowest thermal budgets are all of the quasi-vdW heteroepitaxy type (Zhang \textit{et al.} \cite{Zhang2018} (MOVPE), Dumcenco \textit{et al.} \cite{Dumcenco2015} (CVD) and Chen \textit{et al.} \cite{Chen2018} (CVD)).

Although owing to the general presence of stacking faults, the absence of lattice matching and the challenging unique in-plane epitaxial alignment, the nucleation density of epitaxial 2D chalcogendies acts as a good measure for the intergrain defect density, the nucleation density only, cease to provide the complete picture of defectivity in 2D materials’ epitaxy. Each individual epitaxial system (vdW layer / epi-substrate, growth technique used, growth conditions applied) will always have its individual and characteristic response on (at least one of) these three parameters defining the intergrain defectivity, in addition to intragrain defects such as vacancies and interstitials resulting from impurities and doping. Nevertheless, the presented methodology uncovers the tip of the iceberg of the challenging defect density benchmarking of new epitaxial 2D chalcogenides.

\section{Conclusions}
\label{section:C}

This review represents the emerging need for new materials to further boost the innovation in the semiconductor industry. One of these interesting materials are the vdW materials from which 2D chalcogenides hold lots of promise. The growth approach of epitaxy is presented as one of the most interesting methods to integrate these new materials at large-scale and large-area in a single-crystalline manner with high crystalline quality and low defect density. The epitaxy of 2D chalcogenides is a fast evolving field with many publications following each other, which is attempted to be systematically reviewed in this work. 

Different from conventional epitaxy, the epitaxy of 2D chalcogenides is shown to suffer from flaws like unintentional multiple in-plane epitaxial registries, high densities of stacking faults and the more likely absence of lattice matching, resulting from the weaker coupling between the vdW epi-layers and the growth surfaces. This is emphasised to result in a significant dependence of the nucleation density on the film's defect density, making nucleation density a key parameter for the crystalline quality of the 2D epitaxial growth.

The exposed relation between nucleation density and defect density facilities the generally challenging defect density quantification for these complex materials systems. Therefore, a benchmark on a generalized nucleation density corrected to a surface coverage of 30 \% is performed, that demonstrates the important differences between both quasi-vdW and vdW epitaxy and the impact of the various growth techniques such as MBE, MOVPE and CVD on the epitaxy of 2D chalcogenides. An important trend is highlighted that reveals the general decrease in nucleation density upon increasing growth temperature of the grown 2D chalcogenide, universally for all the discussed growth techniques. The MBE method with its generally moderate growth temperatures systematically yields 2D chalcogenides with the largest nucleation densities, followed by the growth methods of MOVPE and CVD with their typical higher growth temperatures hence lower nucleation (and defect) densities.

Although, every individual 2D chalcogenide epitaxial system has its unique and individual response in the growth characteristics such as grain size, nucleation density, in-plane alignment, strain, stacking faults, intragrain defects, etc., the general observations that are uncovered in this work bring us one step closer to a better identified epitaxial growth process of one the most promising families of future functional materials that are 2D chalcogenides.

\clearpage

\bibliographystyle{ieeetr}
\bibliography{library}

\end{document}